%% file: paper.tex
\documentclass[10pt,sigconf,disable]{acmart}

\usepackage[english]{babel}
\usepackage{blindtext}
\usepackage{algorithm}
\usepackage{algpseudocode}
\usepackage{bytefield}

\renewcommand\footnotetextcopyrightpermission[1]{} 

\setcopyright{none}

\usepackage[colorinlistoftodos,prependcaption]{todonotes}
\presetkeys{todonotes}{inline,linecolor=red,backgroundcolor=red!25,disable}{}

\begin{document}
\title{Adding Forward Erasure Correction to QUIC}

 \author{Fran\c{c}ois Michel, \ \  Quentin De Coninck, \ \  Olivier Bonaventure}
 \affiliation{%
   \institution{UCLouvain}
   \state{Belgium} 
 }
 \email{{francois.michel,quentin.deconinck,olivier.bonaventure}@uclouvain.be}

\begin{abstract}
Initially implemented by Google in the Chrome browser, QUIC gathers a growing interest.
The first stable specification for QUIC v1 is expected by the end
of 2018. It will deliver the same features as TCP+TLS+HTTP/2. 

The flexible design adopted by the IETF for QUIC enables this new protocol to
support a variety of different use cases. In this paper, we revisit the reliable transmission mechanisms
that are included in QUIC. More specifically, we design, implement and evaluate Forward Erasure Correction
extensions to QUIC. Our design supports a generic FEC frame and our implementation includes the 
XOR, Reed-Solomon and Convolutional RLC schemes. We evaluate its performance by applying an experimental
design with a wide range of packet loss conditions. In single-path scenarios, RLC delivers more data than
the two other schemes with short loss bursts. Reed-Solomon outperforms RLC when the bursts are longer. We also apply FEC to Multipath QUIC with a new packet scheduler
that helps to recover more lost packets. 




\end{abstract}

\maketitle

\input{introduction}
\input{background}
\input{design}
\input{implementation}
\input{evaluation}
\input{conclusion}
\newpage 
\bibliographystyle{ACM-Reference-Format}
\bibliography{reference}

\end{document}

%% file: introduction.tex
\section{Introduction}


TCP has been the dominant transport protocol on the Internet for the last decades
but several factors could reduce it in the coming years. Firstly, TCP implementations have been ossified over the years. This ossification has two main causes. First, the Internet contains a variety of middleboxes that make some assumptions on specific TCP variants \cite{honda2011still}. Second, there are many in-kernel implementations of TCP and
any change requires both engineering effort and negotiations within the IETF.
Secondly, the revelations of the utilisation of pervasive monitoring by Edward Snowden have caused the IETF to consider them as an attack \cite{farrell2014rfc}. Since these revelations, we 
observe a fast growth of the utilisation of encrypted protocols. The growing popularity
of TLS and HTTPS is one example of this evolution. 

Google engineers addressed these two
problems by designing the QUIC protocol \cite{langley2017quic}. QUIC started as
an evolution of SPDY \cite{spdy}, a precursor of HTTP/2. In a nutshell, QUIC combines
in a single protocol the mechanisms that are usually found in three different protocols:  TCP, TLS and HTTP. In constrast with TLS/TCP, QUIC encrypts both the payload and most of the protocol headers to prevent both pervasive monitoring and ossification from
middleboxes. Another importance difference between QUIC and TCP is that QUIC runs above
UDP. This implies that QUIC implementations can be included as libraries inside applications. Applications, are regularly updated and recent measurements about the deployment of QUIC show that a dozen different versions of the QUIC protocol were advertised by Google servers during 2017. 

Given the positive results obtained by Google with QUIC \cite{langley2017quic}, the IETF created a QUIC working group in 2016 to standardise a new protocol starting from
Google's design \cite{quic-draft-00}. This group is one of the most active IETF working
groups and the design progresses quickly. The first stable QUIC specification is expected
by the end of the year and more than a dozen interoperable implementations are being
developed. Although the first use case for QUIC will be HTTP/2 \cite{quic-http}, 
the IETF working group will also consider other applications \cite{dnsoquic} once QUIC version 1 becomes stable. 

Internet measurements and traffic predictions \cite{VNI:2017}
show that audio and video streaming applications are now the main traffic sources on the Internet and in mobile networks. The applications do not require strict reliability guarantees and could be served by real-time protocols such as RTP \cite{rtp}. However, given the prevalence of NATs and other types of middleboxes, some of the most popular ones often
rely on HTTP and thus TCP. Given the flexibility of QUIC, it is interesting to 
reconsider the mechanisms that are required to support such applications. 

In this paper, we analyse how QUIC can be extended to support Forward Erasure Correction (FEC) techniques. These techniques transmit redundant code to enable the receiver to recover from packet losses without waiting for retransmissions. FEC techniques have already been used in multicast applications \cite{carle1997survey,perkins1998survey} and even in
TCP \cite{mptcpfec,fmtcp,sundararajan2011network}, but deploying the latter remains difficult. This paper is organised as follows. We first describe the characteristics of QUIC and the concepts behind Forward Erasure Correction. We then define the design and implementation details of QUIC-FEC, our extension enabling the use of FEC-protected unreliable transfer with QUIC. We finally assess the performances of our extension, compare our different FEC Schemes and study the benefits of a multipath communication through experiments using a wide range of loss configurations. 

%% file: background.tex
\section{QUIC and Forward Erasure Correction}
The QUIC protocol is a newly introduced transport protocol built atop UDP already representing more than 7\% of the total Internet traffic~\cite{langley2017quic}.
A key element of the QUIC design is its extensibility.
First, it includes a version negotiation procedure allowing both hosts to agree on a specific QUIC flavour.
Second, since all data and control information is encrypted, QUIC is most robust towards attacks and middleboxes, known to slow down the deployment of protocol extensions ~\cite{tcpmiddlebox, deployingmptcp, honda2011still}.
Third, hosts can exchange connection-specific parameters during the handshake, called \emph{transport parameters}, to tune the protocol to the traffic it carries.
Among other main QUIC features, there are 0-RTT connection establishment saving the TCP 3-way handshake and stream multiplexing useful for HTTP/2.
Using multiple QUIC streams allows hosts to carry independent data over a single connection without experiencing TCP head-of-line blocking~\cite{quicexperiments}. 


A QUIC packet is 
divided into two parts: the header, which is authenticated, and the payload, which is both authenticated and encrypted. The QUIC payload can be seen as a container for one or multiple QUIC frames, which are all independently handled by the receiver. There are tens of types of frames. Among them, the frames containing the application data are the \textit{STREAM frames}, while the frames acknowledging the received packets are the \textit{ACK frames}. 
QUIC integrates flow control, where the receive window is advertised with \emph{WINDOW\_UPDATE frames}.
Using the frame abstraction allows QUIC to be more modular and extensible, by defining new protocol behaviours for handling new frames and letting the remaining of the protocol unchanged.

There are two main designs of QUIC. The original design of QUIC, currently called \textit{Google-QUIC} (or \textit{gQUIC}) \cite{gquic} was designed and deployed by Google. Convinced by the results obtained with \textit{gQUIC}, the IETF has created
a working group to standardise \textit{IETF-QUIC}\cite{quic-draft-11}. While the Google-QUIC implementations are supposed to transition towards IETF-QUIC once
the standardised protocol matures, the QUIC implementation used in this paper has not yet completed this shift. Furthermore, while IETF-QUIC is still regularly evolving, Google-QUIC is currently more mature and deployed at a larger scale on the Google servers \cite{takinglonglookatquic}. For this reason, we focus on Google-QUIC in this study. Carlucci \textit{et al.} \cite{quicexperiments} show that Google-QUIC can outperform HTTP/1.1 over TCP with TLS and reduce the page load times when there are no packet losses. They also show that QUIC can outperform SPDY \cite{spdy} but not necessarily HTTP/1.1 over TCP with TLS when there are random losses.  Megyesi \textit{et al.} \cite{howquicisquic} also show in their analysis that although QUIC can reduce the page load times, the benefit of QUIC over other protocols is not systematic and often depends on the network conditions.


\subsection{Forward Erasure Correction}

Most transport protocols rely on variants of Automatic Repeat-Request (ARQ) techniques to cope with transmission errors and losses. This is not the only possible solution. Over the last decades, researchers have explored a variety of techniques that transmit redundant data to enable the receiver to recover from errors and losses without having to wait for retransmissions. Some of the proposed techniques were tuned for specific to link layer technologies \cite{biersack1993performance,katti2008xors} or targeted for specific applications \cite{carle1997survey,perkins1998survey}. Some of them have even been proposed for TCP \cite{sundararajan2011network,fmtcp,mptcpfec}. The IETF also considers these techniques within the RMT and FECFRAME working groups and the IRTF NWCRG \cite{Adamson_Taxonomy:2018}. 

Given that most link layer technologies include error detection codes, most of the proposed solutions focus on Forward Erasure Correction (FEC), i.e., the ability of sending redundancy (Repair Symbols) along with the data that will help to recover the data packets (Source Symbols) that have been lost. This allows recovering from packets losses without packet retransmission. We use the word \textit{FEC Scheme} to refer to the way to handle Source and Repair Symbols and generate the redundancy using an erasure correcting code.



While FEC was originally part of the QUIC protocol~\cite{quic-draft-00}, it has rapidly been dropped due to negative experiments~\cite{takinglonglookatquic}.
However, the considered FEC scheme only allowed recovering single packet losses using a XOR code, leading to poor recovery capabilities with often correlated losses~\cite{quicfecprague}.
A wide range of codes more adapted to those network conditions have been proposed in the literature. In this work, we consider both block and convolutional codes. 

\subsubsection{Systematic block codes}

A $(n, k)$ block code  is an error correcting code that takes a block A of $k$ M-ary symbols and maps it to a block B of $n$ M-ary symbols such that $k < n$. In a \textit{systematic} block code, the input codeword is present in the output codeword unchanged. The $n-k$ remaining symbols of the output codeword are the \textit{Repair Symbols}. The code rate of a $(n, k)$ block code is equal to $\frac{k}{n}$.

An optimal $(n, k)$ block code can stand the loss of $n-k$ symbols. An example of optimal block code is the well-known Reed-Solomon code \cite{reedsolomoncodes}. When working with a stream of symbols, the sequence of symbols is split into several blocks that will be independently protected. Galanos \textit{et al.} \cite{rtp-reedsolomon} provide an internet draft for the use of the Reed-Solomon block codes with the RTP protocol \cite{rtp}. The Reed-Solomon block code is currently suggested among the reference coding schemes to be used with the QUIC protocol, according to the standardisation effort of the IRTF \cite{coding-for-quic}.
\todo{OB/ list some protocols that use this technique}

\subsubsection{Systematic convolutional codes}

A \textit{convolutional} code (a.k.a. \textit{sliding window} code) is an error correcting code based on the incremental generation of the Repair Symbols by applying a sliding window computation on the sequence of Source Symbols to be sent. Instead of considering blocks of Source Symbols, we see the Source Symbols as a stream and perform the coding incrementally. This means that one Source Symbol can be used multiple times to compute different codewords. The window has a length $L$, performs shifts of $k$ Source Symbols at each iteration and outputs a codeword of $n$ symbols for each iteration, implying that the code rate is $\frac{k}{n}$. We denote such a code as a $(n,k,L)$ convolutional code. 

\begin{figure}[t]
	\centering
	\includegraphics[scale=0.4]{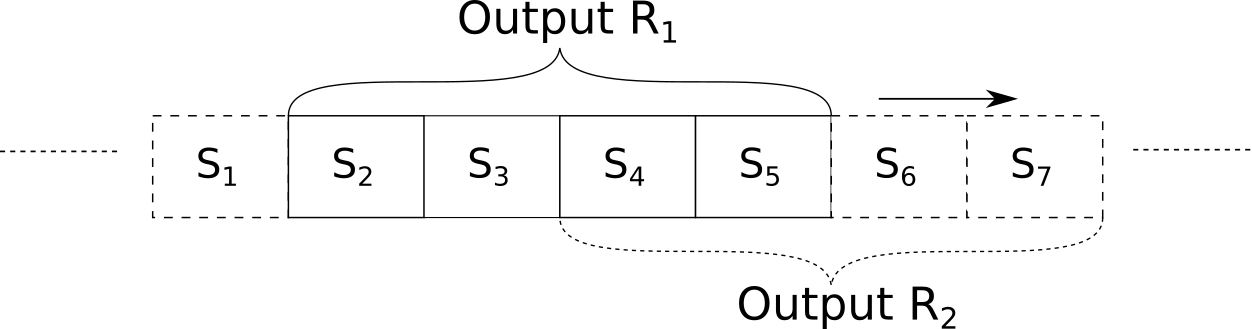}
	\caption{Example of convolutional code.}
	\label{convo_code_example}
\end{figure}

Figure \ref{convo_code_example} shows an example of a $(3, 2, 4)$ convolutional code: the window of size 4 slides with a step of 2 on the symbol stream and outputs one Repair Symbol at each step. Note that the Source Symbols $S_4$ and $S_5$ are used to compute both $R_1$ and $R_2$. An example of convolutional code is the convolutional Random Linear Code (RLC) \cite{rlc-fecframe}. For each window position, the code generates $n-k$ linear equations (one for each Repair Symbol) whose constant term is the Repair Symbol and whose variables are the Source Symbols present in the encoding window. The coefficients of the equations are randomly generated using a PRNG, reducing the probability of having co-linear equations. 




Roca \textit{et al.} compare block and convolutional codes using Reed-Solomon and RLC to represent both families of codes \cite{roca2017less}. They show that while the Reed-Solomon block codes provide a higher encoding speed, RLC allows to recover the packets with a reduced latency compared to Reed-Solomon. We leverage this property with our QUIC extension.

\subsubsection{Packet-level coding}

In a packet-based transport protocol like UDP, the packet reception outcome is binary. Either the packet is received or it is not received at all. We can thus consider entire packets as symbols, instead of protecting an arbitrary number of bits in the packets. This is called \textit{packet-level coding}. A systematic $(n, k)$ block code using packet-level coding considers a block of $n$ packets in which $k$ packets are the $k$ \textit{Source Symbols} and the $n-k$ remaining packets are the $n-k$ \textit{Repair Symbols}. Using packet-level coding allows us to focus on \textit{erasures} instead of errors. Indeed, as QUIC already performs an integrity check on the packets, a packet containing an error will be dropped and considered as a symbol erasure.

%
%



%% file: design.tex
\section{Integrating FEC into QUIC}



Both Google-QUIC and IETF-QUIC provide a fully reliable data transfer. This is the service required to support HTTP/2 which is the motivating use case for QUIC. However, there are applications that have strict expiration deadlines. 
Such applications include live streaming cases. 
An unreliable data transfer can be useful in these use-cases.
Indeed, when the data expiration deadline is over, the retransmission of the data does not make sense anymore and will imply a waste of time and bandwidth. We thus introduce the \textit{Unreliable Streams} in QUIC. Those streams ensure that the data are provided in order, 
potentially with 
gaps in the data when the stream is processed. We introduce the QUIC \textit{Unreliable Stream frame} that carries the data of an Unreliable Stream. Unreliable streams are explicitly advertised as such to the applications.

While packets might be lost, they can be recovered by the reception of Repair Symbols.
In this section, we propose a generic Forward Erasure Correction extension of QUIC. It supports different FEC Schemes in a transparent manner. The application can select the FEC Scheme that suits its needs and recover from losses without waiting for  retransmissions. We first describe our choice of Source and Repair Symbols representations. We then present our FEC Framework enabling the usage of different FEC Schemes and simplifying their implementation. 

\subsection{Representation of Source Symbols}

In order to use Forward Erasure Correction, one needs to define the Source Symbols on which the Forward Erasure Correction will be applied. QUIC offers frames and packets as data containers.
 
\subsubsection{QUIC packets as Source Symbols}

We use unencrypted QUIC packets as Source Symbols. While using QUIC stream frames is an interesting candidate for being a Source Symbol, using a packet instead 
guarantees that the loss of a packet exactly corresponds to the loss of one Source Symbol. To perform the coding and decoding with packets of different sizes, some packets may need to be padded with zeros. There is however no need to encode the packet length to recover the packet correctly. Indeed, padding is naturally understood by the QUIC protocol thanks to the padding frame which has no content and frame type \texttt{0x0}. The only difference between a packet and its recovered version is that the latter will potentially contain several padding frames at its end. Nevertheless, this requires a \textit{Data Length} field in every stream frame of a FEC-protected packet. 

Using unencrypted QUIC packets as Source Symbols implies that the Repair Symbols must be sent encrypted and authenticated. 
This prevents both adversarial modifications of the Repair Symbols and recovering of the Source Symbols by eavesdropping the Repair Symbols.


Encrypting the Repair Symbols could have been avoided by considering encrypted packets as Source Symbols. However, the padding is not naturally handled anymore and the packet length should be added in its clear-text header to avoid encryption and authentication problems, adding a high complexity to the solution without concrete benefit.

\subsubsection{Distinguishing Source Symbols from regular packets}

Although Forward Erasure Correction allows recovering lost packets without waiting for retransmissions, it consumes more bandwidth than a regular, non-FEC-protected transmission. In order to avoid spending additional bandwidth when it is not needed, our design must allow defining which QUIC packets should be considered as FEC Source Symbols. Our design allows explicitly distinguishing the packets being FEC-protected from the packets that are not.


The Public Header of Google-QUIC has been slightly adapted to support this distinction. Figure~\ref{quic_public_header_modified} shows the modifications. We use the unused \emph{F} flag from the Google-QUIC Public Header to indicate whether a packet is FEC-protected or not.
If the flag is set, the 
\emph{Source FEC Payload ID} field is appended to the header.
Such design, inspired by FECFRAME~\cite{fecframe}, offers a 32-bits field that can be used by the underlying FEC Scheme.
The field is opaque for the transport protocol. 


\begin{figure}
\centering
\resizebox{7cm}{!}{
\begin{bytefield}[bitwidth=0.85em]{32}
\bitbox[ltr]{1}{\textbf{F}} \bitbox[ltr]{7}{Flags (7)} \\
\bitbox[ltr]{32}{Connection ID (64) (optional)} \\
\bitbox[ltr]{32}{Packet Number (8, 16, 32 or 48) (variable length)} \\
\bitbox[lrtb]{32}{\textbf{Source FEC Payload ID (32) (if F set)}}\\
\bitbox[lrb]{32}{Encrypted payload ...} \\
\end{bytefield}
}
\caption{Modified Google-QUIC Public Header.}
\label{quic_public_header_modified}
\end{figure}


\subsection{Transmitting Repair Symbols}

The Repair Symbols must be distinguished from application data payload in the sense that it should not be transferred to the application upon reception. The Repair Symbols are indeed generated by and for the FEC Scheme used by the transport protocol. 

We consider three ways to send Repair Symbols to the QUIC receiver: $(i)$ using a dedicated packet type, $(ii)$ using a dedicated stream or $(iii)$ a dedicated frame. Using a dedicated packet type seems to be the solution with the smallest overhead, only needing a packet header. However, we need to define a new packet type and it could be a painful process in some QUIC implementations to introduce this new change. Using a dedicated stream has several drawbacks: the stream abstraction implies an in-order data delivery while the Repair Symbols could be processed in any order and should be processed as soon as possible. Furthermore, reserving a stream ID for FEC would add an additional corner case for FEC, increasing the complexity of QUIC implementations with special streams~\cite{quicdtls}. 
For these reasons, we use a dedicated frame, the \textit{FEC frame}. Using a new frame type is a natural and easy way to extend the QUIC protocol with new behaviours.
Furthermore, as QUIC frames sent after the cryptographic handshake are always encrypted and authenticated, the FEC frame ensures the Repair Symbol properties required by our previous choice of taking unencrypted packets as Source Symbols.

The FEC frame transports Repair Symbols. Each Repair Symbol is associated with a 64-bits \textit{Repair FEC Payload ID}. The Repair FEC Payload ID is an opaque field for the protocol. It is used by the FEC Scheme to identify the Repair Symbols and communicate information about the encoding and decoding procedures to the receiver-side FEC Scheme.

It is preferable to minimize the number of packets that carry a single Repair Symbol to ensure proper reception of the whole symbol.
However, a FEC Scheme might generate Repair Symbols that cannot fit in a single QUIC packet.
Our design allows seamlessly splitting Repair Symbols into several FEC frames.




\subsection{The FEC Framework}

The IETF has already developed solutions to add Forward Error
Correction codes to several protocols. The most recent solution
is the FECFRAME framework \cite{fecframe} which has notably been
applied to RTP and supports different FEC schemes 
\cite{reedsolomonfecframe,rfc6816}. 



Although there exists a wide variety of different FEC Schemes, we focus in this paper on two main categories: FEC Schemes using block codes and FEC Schemes using convolutional codes. 
Inspired by FECFRAME \cite{fecframe}, we define a FEC Framework implementing the common behaviour of these FEC Schemes in order to simplify their implementation. It provides a structure for the Source and Repair FEC Payload ID's that are opaque to the underlying protocol.

\begin{figure}
\includegraphics[scale=0.6]{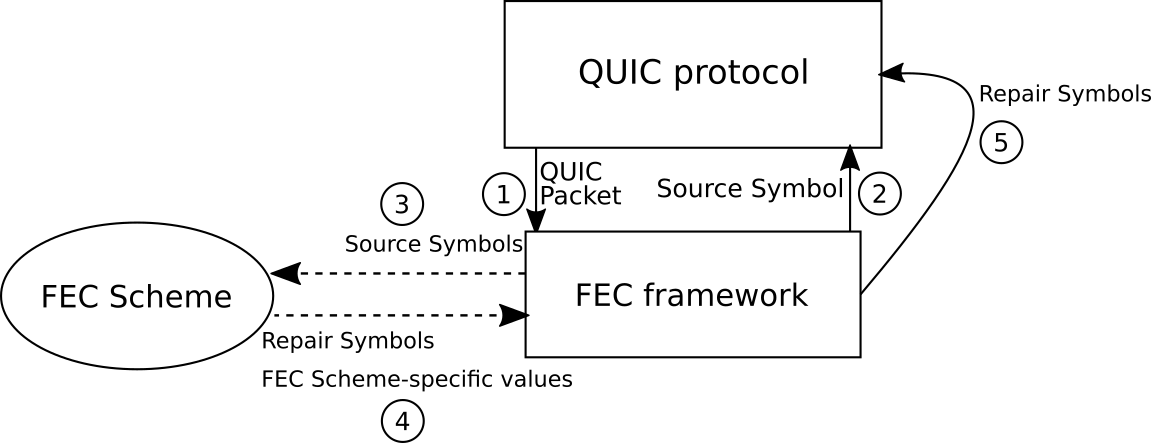}
\caption{Interactions between the protocol, the FEC Framework and the FEC Scheme.}
\label{fig_fec_framework}
\end{figure}

Figure \ref{fig_fec_framework} shows the interactions between the QUIC protocol, the FEC Framework and the FEC Scheme. The protocol first sends to the FEC Framework the QUIC packets that must be protected (1). The FEC Framework then transforms this packet into a Source Symbol (i.e., a QUIC packet with the header shown in Figure \ref{quic_public_header_modified}) and returns it to the protocol in order to send it as soon as possible (2). When necessary, the FEC Framework sends its Source Symbols to the FEC Scheme to generate Repair Symbols (3). For each Repair Symbol, the FEC Scheme also generates \textit{FEC Scheme-specific values} related to the encoding procedure and gives them to the FEC Framework along with the Repair Symbols (4). The FEC Framework then generates the Repair FEC Payload ID's, attaches it to the Repair Symbols and passes the Repair Symbols to the QUIC protocol for them to be sent (5). Once the protocol receives the Repair Symbols, they are sent to the receiver through the previously mentioned FEC Frames. At the receiver-side, the received Source Symbols can be processed immediately. The Repair Symbols are reconstituted from the FEC Frames and passed to the underlying FEC Scheme to recover the lost Source Symbols.

\subsubsection{Negotiating a FEC Scheme}

There are different FEC schemes and a streaming application has different needs than an a request-response application that wants to cope with tail loss. The application that uses QUIC-FEC should be
able to negotiate the FEC Schemes that are used to encode and decode the Repair Symbols in each direction. We carry out this negotiation during the QUIC cryptographic handshake using the transport parameters. One FEC scheme is selected in each direction: the $client \rightarrow server$ FEC Scheme can differ from the $server \rightarrow client$ one.
Having different FEC schemes makes sense in a configuration where the client has lower computational capabilities than the server.
In such a configuration, it would be useful to choose a $server \rightarrow client$ FEC Scheme whose decoding procedure has a low computational complexity and a potentially different $client \rightarrow server$ FEC Scheme whose encoding procedure has a low complexity on the client.

Such interfacing also brings interest from the IETF, where the network coding research group currently works on an Internet Draft~\cite{coding-for-quic}.
While core ideas are similar, this draft is still currently a work in progress, as it only provide guidelines.
We believe our work will benefit to the standardisation of such interfacing.

%% file: implementation.tex
\section{Implementation}

QUIC-FEC includes both our proposed Forward Erasure Correction and the Unreliable Streams. We implement QUIC-FEC on top of \texttt{quic-go} \cite{quic-go}. In total, we added $\sim$ 5000 
lines of code. $\sim$ 1500 lines have been added to already existing files and $\sim$ 3500 lines have been added to dedicated files. A patch of $\sim$ 200 KB has been generated with the Linux \texttt{diff} tool. 
The main reason why we opted for this implementation as the starting point for our work is to  be able to also test with the
Multipath QUIC version that was released as
patches to this implementation~\cite{mpquic-design,mpquic}. 

\subsection{XOR FEC Scheme}

The  principle of the XOR FEC Scheme is quite simple: the Source Symbols are simply XORed with each other to generate a Repair Symbol. 
Although this scheme is easy to implement and compute, it can only recover the loss of one Repair Symbol. Experiments carried out
by Google show that this is insufficient on the Internet \cite{Swett_FEC:2016} because losses can occur in bursts. This is why our implementation uses interleaving to recover from burst losses with the XOR FEC Scheme. 
Sending successive packets in different FEC Blocks enables simple FEC Schemes such as XOR to better handle burst loss at the expense of delay.

\subsection{Reed-Solomon FEC Scheme}

In addition to the XOR FEC Scheme, our implementation also supports a Reed-Solomon FEC Scheme to generate the redundancy. We reuse the \texttt{klauspost/reedsolomon} Reed-Solomon implementation \cite{Post_RS}. This implementation provides a systematic block code that can take up to 256 Source Symbols as input and generate up to 256 Repair Symbols. Compared to the XOR FEC Scheme, the Reed-Solomon can thus generate multiple Repair Symbols per Source Block, enabling the FEC Scheme to naturally handle burst losses.

\subsection{Convolutional Random Linear Code FEC Scheme} 

As convolutional FEC Schemes provide different properties from Block FEC Schemes, our implementation also enables their use through the Convolutional RLC FEC Scheme. Our implementation relies on the \texttt{alex-ant/gomath} linear equations system solver \cite{Antonov_go-math}. 
Using Gaussian Elimination, recovering from lost Source Symbols is costly from a CPU viewpoint. In order to avoid performing too many computations, we only trigger the gaussian elimination process if the equation system is square or if it contains a square subsystem. Our implementation is highly inspired from the \textit{FECFRAME} RLC FEC Scheme draft \cite{rlc-fecframe}. The sender FEC Scheme sends a \textit{FEC Scheme-Specific} value to the receiver-side FEC Scheme. This value is composed of the 16-bits seed used to generate the random coefficients and the \textit{Density Threshold} representing the proportion of non-zero coefficients in the equations. Our implementation uses the Park \& Miller PRNG \cite{parkmiller} to generate the random equation coefficients. \footnote{As of writing this article, the fifth version of the draft has been released and advises to change the PRNG \cite{rlc-fecframe-05}. This will be done in the near future.}


%% file: evaluation.tex
\section{Methodology}

Experiments have been
performed to assess the performances of our implementation and analyse the benefits of Forward Erasure Correction with real-time applications running over QUIC. 
We describe in this section our methodology.


We used network emulation with the Mininet tool \cite{mininet} to evaluate the performance of different FEC schemes in \texttt{quic-go}. The
main benefit of using emulation is that it runs with real code and
not a simplified protocol model. On the other hand, network emulation
has limitations when the throughput is high and/or a large number of nodes must be emulated on a host with a small number of CPUs.
We mitigate those problems by using sources that send at 
low throughput and map each node to its own CPU core. 

\subsection{Loss model}

A wide range of loss models have been proposed in the literature. Some of them consider either uniform random losses, short burst sizes or pre-defined uniform loss rate variation to represent loss bursts, such as those of Carlucci et al. \cite{quicexperiments}, Cui et al. \cite{fmtcp}, Dong et al. \cite{lamps} and Ferlin et al. \cite{mptcpfec} (which considers an average loss burst length of 2). Others consider a burst loss model with potentially long loss bursts like Roca et al. \cite{roca2017less}, Tournoux et al. \cite{tetrys} and Badr et al. \cite{badr2013streaming}. While small burst lengths or uniform losses can already give an idea on the efficiency of a solution, we advocate for looking at longer burst lengths in order to evaluate our solution. We thus consider the use of the Gilbert-Elliott model \cite{elliottmodel}, allowing to represent bursts of lost packets. 

\begin{figure}
\centering
\includegraphics[scale=0.15]{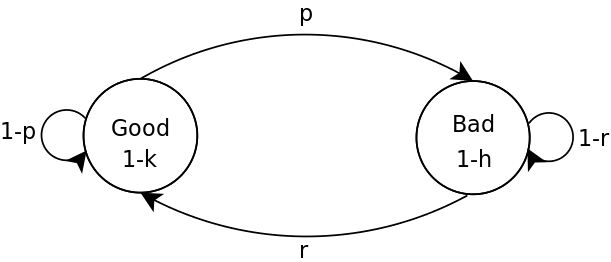}
\caption{The Gilbert-Elliott loss model.}
\label{fig_gemodel}
\end{figure}

The \textit{Gilbert-Elliott} model is shown in Figure \ref{fig_gemodel}. It is a two-states Markov model. The two states are the \textit{Good} and \textit{Bad} states. In the \textit{Good} state, a packet is delivered with a probability $k$. In the \textit{Bad} state, a packet is delivered with a probability \textit{h}. \textit{p} denotes the probability of transition from the \textit{Good} to the \textit{Bad} state, while \textit{r} denotes the probability of transition from the \textit{Bad} to the \textit{Good} state. As Mininet only provides a uniform loss model, we patched it to use the Gilbert-Elliott loss model provided by the \texttt{netem} Linux tool \cite{netem}.

\subsection{Experimental design}

We use the \textit{experimental design} approach to perform our experiments \cite{fisher1949design}. 
This methodology consists in defining ranges of possible values for each parameter and performing a series of experiments with random values chosen within these ranges. This sub-samples the ranges of values and gives a global overview of the possible values taken by all the parameters. In addition to providing a general confidence concerning the performances of the tested implementation, it mitigates the bias in the parameters selection by the experimenter. This bias could have led to selecting unrealistic parameters values or avoiding special cases in which the implementation would perform poorly. Using the experimental design approach also permits 
to explore edge cases that would not have been explored otherwise.
We use the WSP algorithm \cite{wsp} to sample broadly the space of parameters with a reasonable amount of experiments. Unless otherwise specified, we run the experiments with 120 different combinations of parameters.
Each configuration is ran 3 times and the median run is considered.
Table \ref{table_parameters_range} shows the parameters ranges chosen for our experiments. We do not limit the bandwidth on the links as we want to study the loss recovery capabilities that can be achieved with our solutions.  The one-way delay (OWD) ranges from 0 to 100 milliseconds, leading to a maximum round-trip-time of 200 milliseconds. This is not completely in line with the measurement-based study of Zhang \textit{et al.} \cite{zhang2006measurement}. In this study, although the majority of the measured round-trip-times is between 0 and 200 milliseconds, some of them can reach up to 400 milliseconds. We further perform in this section a univariate analysis targeted on the delay to show whether high delays have a negative influence on our solutions.

When it is not otherwise specified, the chosen FEC Scheme is the Reed-Solomon FEC Scheme and the level of redundancy is set to $(30, 20)$. This ensures a code rate of $\frac{2}{3}$ and a burst recovery capability of 10 symbols per block. By looking at the possible values for the $r$ parameter in Table \ref{table_parameters_range}, we can see that we will often be in configurations where the average burst length is longer than 10 packets, possibly leading to poor performances of our solution. This is done on purpose for multiple reasons. First, with experimental design, we want to assess the global performances of our solution, even for the cases that are not in favour of it. Second, when using the Gilbert-Elliott model, an average burst of $n$ packets in the \textit{Bad} state does not imply that the average number of packets lost in the burst will be $n$ too. We thus want a reasonable margin to reach the undesirable cases even with a Gilbert-Elliott model. Finally, having an average burst length longer than the recovery capabilities of our solution will allow us to take a look at the benefits of multipath in terms of symbols recovery. Although the majority of experiments focus on these parameter values, we also explore other values for some experiments.

\begin{table}
\centering
\begin{tabular}{|c|ccccc|}
\hline
Parameter & $p$ & $r$ & $k$ & $h$ & OWD (ms)\\
\hline
Smallest & 0 & 0.025 & 0.97 & 0 & 0\\
Highest & 0.01 & 0.5 & 1 & 0.4 & 100\\
\hline
\end{tabular}
\caption{Experimental design parameter ranges.}
\label{table_parameters_range}
\end{table}

\subsection{Traffic sources}

Our experiment focuses on a real-time data transfer, inspired from live video streaming applications. We ignore frame compression and audio to ease the implementation and the interpretation of the tests. The sender regularly sends application messages composed of several QUIC packets at a constant bit rate. 
In our experiments, we use 30 messages per second, with 8 QUIC packets of 1000 bytes per message, which implies that the application messages are spaced by $\sim 33$ milliseconds with a bit rate close to 2~Mbps. 
Each experiment runs during 25 seconds, in order to encounter losses even with low loss probabilities.

The receiver reads the application messages at the same rate and reports statistics about the user experience. A message is considered well received when all its parts have been delivered. It is considered as corrupted if one or more parts are missing. When the receiver cannot read a message on time (i.e., it has to wait more than 33 milliseconds before reading a message) either because it is received too late or corrupted, the receiver considers that some re-buffering occurred. The duration of the re-buffering depends on the number of successive messages that could not be delivered on time.

In order to directly focus on the recovery capabilities of our solutions, we have disabled the congestion control for all experiments. Indeed, a pacing-based congestion control as the one used by QUIC \cite{cubic} is inappropriate for real-time communications \cite{rtpcongestion} as it unnecessarily delays packets. Real-time applications adapt to such congestion by adapting their sending rate (e.g., by compressing its video frames for a video application or by reducing the quality of the images), but this behaviour is outside the scope of this study.


In order for the FEC Scheme to be able to recover from packet losses, we introduce a fixed playback buffer on the receiver. This buffer allows the FEC Scheme to receive enough Source and Repair Symbols to recover the lost Source Symbols. In our experiments, we use a default playback buffer of 100 milliseconds. We explore other playback buffers in dedicated experiments.

\subsection{Reported metrics}
For each test, different metrics are measured and reported.


\paragraph{Fraction of correctly received application data}
This metric reports the fraction of the data sent that has been delivered to the receiving application. It only considers the amount of data received though the QUIC Unreliable Stream used by the application: it does not consider the Repair Symbols, packet headers or other QUIC frames.  

\paragraph{Total re-buffering time}
This metric is a good indicator of the quality of experience perceived by the end-user as it represents the total wasted time during the connection. With a rate of $x$ application messages per second, a lost or corrupted application message will imply a re-buffering of $\frac{1}{x}$ seconds (33 milliseconds in our experiments). If the application only receives a part of an application message, this part is still taken into account in the amount of data received but a re-buffering is reported in that case, as the application message has been corrupted. Having a scenario where the majority of the sent data has been received and a high total re-buffering time has been reported indicates that many application messages have been corrupted with only a small part of missing data.


\section{Single path Experiments}
We first assess the performance of our solution in a single-path configuration.

\begin{figure*}
	\hspace{-0.8cm}
	\begin{minipage}[t]{.34\linewidth}
		\captionsetup{width=.95\linewidth}
		\includegraphics[scale=0.19]{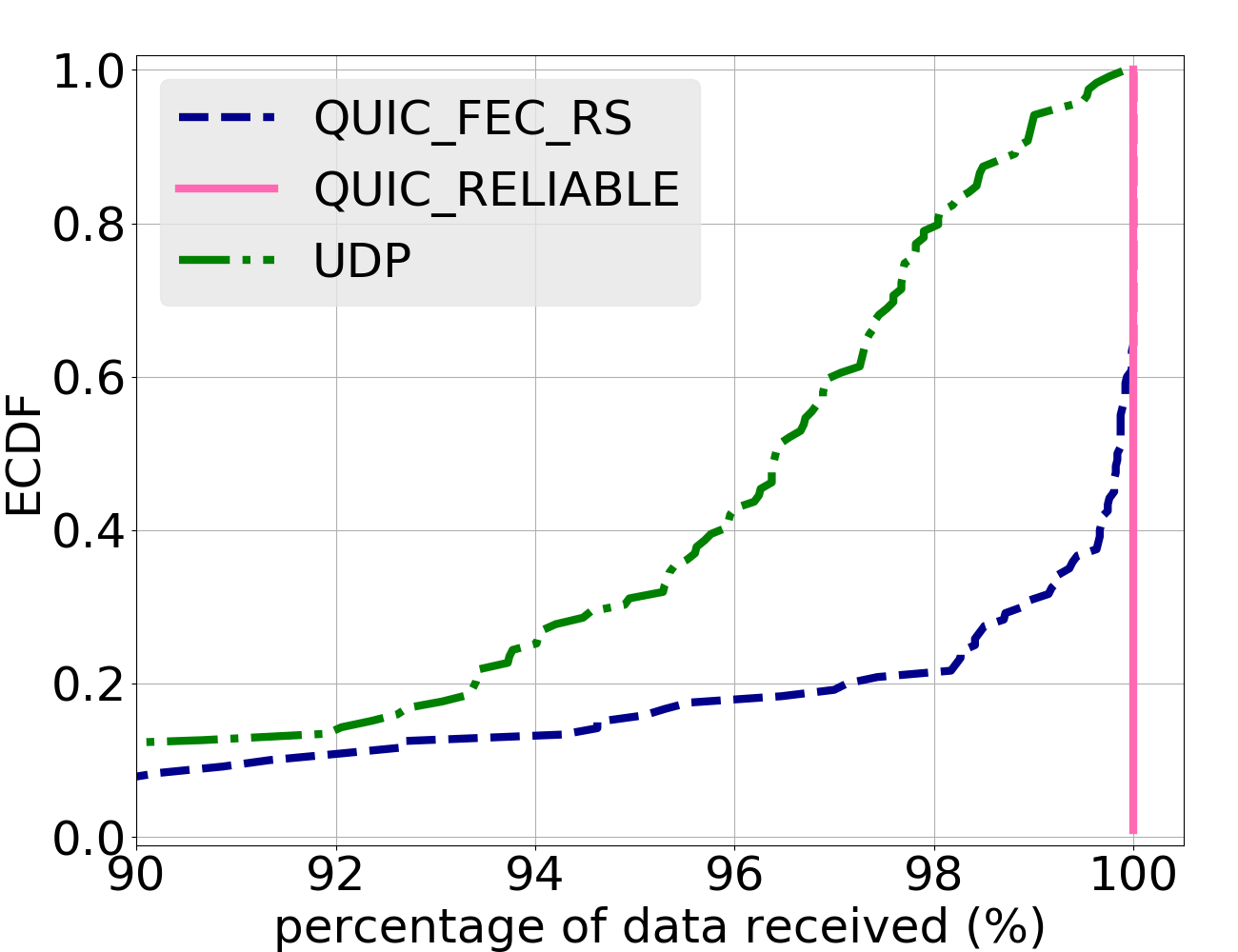}
		\caption{UDP VS Reed-Solomon VS Reliable QUIC: amount of application data received.}
		\label{fig_quic_udp_quicfec_data}
	\end{minipage}
	\begin{minipage}[t]{.34\linewidth}
		\captionsetup{width=.95\linewidth}
		\includegraphics[scale=0.19]{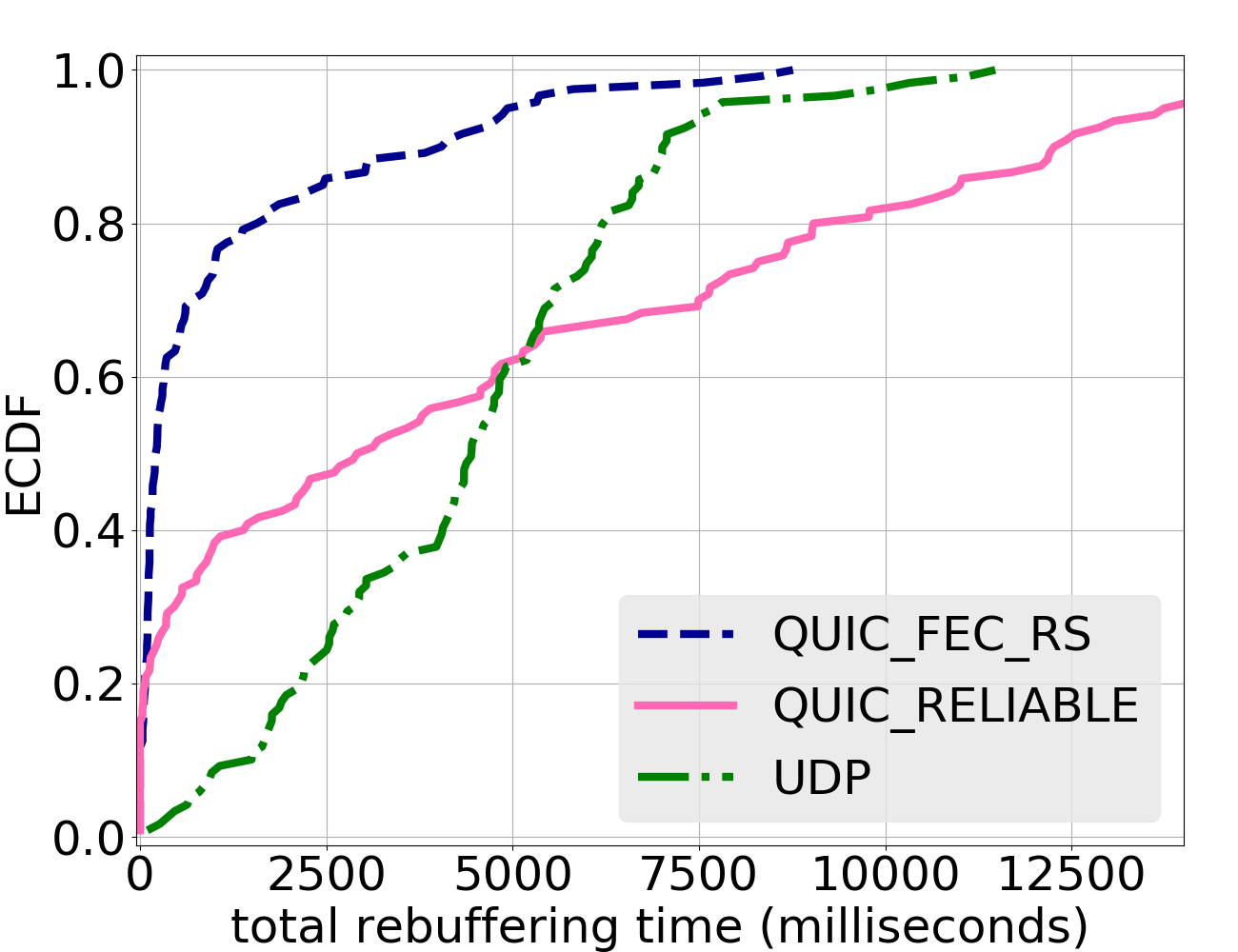}
		\caption{UDP VS Reed-Solomon VS Reliable QUIC: total re-buffering time.}
		\label{fig_quic_udp_quicfec_rebuffering}
	\end{minipage}
	\begin{minipage}[t]{.34\linewidth}
		\captionsetup{width=.95\linewidth}
		\includegraphics[scale=0.19]{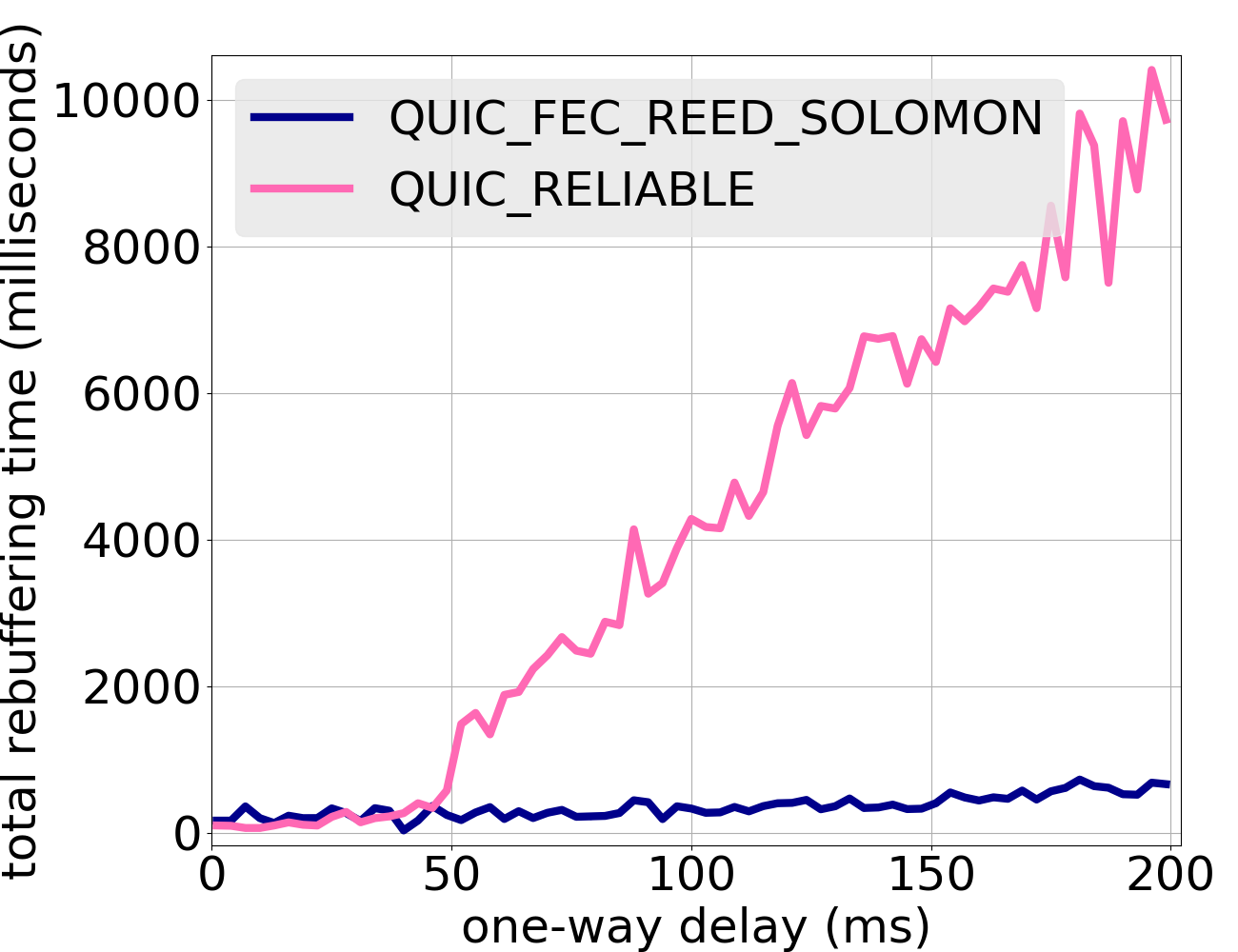}
		\caption{QUIC-FEC VS reliable QUIC: total rebuffering time w.r.t. the one-way delay.}
		\label{fig_quic_fec_quic_univariate_delay}
	\end{minipage}
\end{figure*}

\subsection{Comparing QUIC-FEC to UDP and QUIC}

Figure \ref{fig_quic_udp_quicfec_data} compares the Experimental Cumulative Distribution Function (ECDF) of the amount of application data received for QUIC, QUIC-FEC and UDP.  The curves are computed from the median experiment over three runs of 25 seconds on 120 configurations, each using a different combination of parameters in the ranges defined in Table \ref{table_parameters_range}. As expected, the reliable QUIC received all the sent application data: all the points of the reliable QUIC curve are on the 100\% line. QUIC-FEC receives a lot more data than UDP: QUIC-FEC receives all the sent application data in more than 35\% of the cases while UDP never receives all the sent data (no point of the UDP curve lies on the 100\% line). This shows the advantage of sending Repair Symbols along with the data. However, we can see that QUIC-FEC recovers less data than reliable QUIC. Figure \ref{fig_quic_udp_quicfec_rebuffering} shows the ECDF of the total re-buffering time for the three solutions. Although the reliable QUIC solution receives all the data in all our experiments, we can see that it generally encounters a higher total re-buffering time than QUIC-FEC: the reliable QUIC encounters a total rebuffering time of more than 2500 milliseconds in more than 50\% of our experiments while QUIC-FEC encounters a similar total rebuffering time in less than 10\% of our experiments. 
Furthermore, the curves of reliable QUIC and UDP cross at some point in Figure \ref{fig_quic_udp_quicfec_rebuffering}. After taking a closer look at our results, the samples on the reliable QUIC curve that are to the right of the UDP curve concern parameters configurations with a high delay. A high delay implies a high retransmission timeout for the QUIC sender. In that case, the playback buffer of 100 milliseconds used by the application may not be sufficient to receive the retransmission of the lost data on time. On the other side, the UDP receiver simply skips the lost frames without blocking, which gives a smaller re-buffering compared to reliable QUIC when the delay is high. QUIC-FEC tries to recover the lost data on time with FEC and skipping the data that could not be recovered.  This shows the advantage of an unreliable data transfer combined with Forward Erasure Correction compared to a retransmission-based reliable data transfer.

\subsubsection{The impact of the delay on a retransmission-based protocol}

Thanks to the experimental design, we have seen in the results of the previous experience that the reliable QUIC solution performs badly when the one-way delay (OWD) is large. In order to better analyse the impact of the delay on reliable QUIC and QUIC-FEC, we perform a univariate experiment. All parameters have been fixed except the delay. The value of the parameters are shown on Table \ref{table_parameters_range_univariate_delay}.

\begin{table}
\centering
\begin{tabular}{|c|ccccc|}
\hline
Parameter & $p$ & $r$ & $k$ & $h$ & OWD (ms) \\
\hline
Value & 0.005 & 0.25 & 0.98 & 0.05 & 0 to 200 \\
\hline
\end{tabular}
\caption{Parameter ranges for the univariate analysis on the delay.}
\label{table_parameters_range_univariate_delay}
\end{table}

Figure \ref{fig_quic_fec_quic_univariate_delay} shows the total re-buffering time of reliable QUIC and QUIC-FEC in function of the one-way delay. We can see that when the delay is low, both reliable QUIC and QUIC-FEC have a low re-buffering time. When the delay becomes larger than 45 milliseconds, the total re-buffering time of reliable QUIC increases a lot. \texttt{quic-go} considers by default that a packet has been lost after waiting $\frac{9}{8}*\operatorname{RTT}$ since its transmission without receiving an acknowledgement for it,
which was in line with the IETF recommended values until December 2017~\cite{quic-rec-draft-07}.
This means that the \texttt{quic-go} receiver will have to wait an additional time of $\frac{9}{8}*\operatorname{RTT}$ to receive a the retransmission of a lost stream frame. With a delay higher than 45 milliseconds, the receiver will thus have to wait more than 100 milliseconds before receiving the retransmission of the lost data. With a playback buffer of 100 milliseconds, it will imply rebufferings at the receiver side. We can see that QUIC-FEC does not suffer from this problem as it does not rely on retransmissions to recover lost data and skips the lost data that could not be recovered. It is thus robust against long delays. There is however a slight increase in the total re-buffering time for QUIC-FEC with high delays. This is due to the QUIC flow control. With long delays, the sender will be blocked at the beginning of the connection, as the receive window of the receiver is full. The \emph{WINDOW\_UPDATE} frames that extend the receive window will take more time to arrive to the sender with high delays, which increases the time during which the sender cannot send anything. This problem can be mitigated by increasing the initial receive window of the QUIC receiver depending on the expected delay. However, it will still remain vulnerable to longer delays. The only way to avoid this problem completely would be to disable the flow control, which is undesirable as it protects the receiver from being overwhelmed by the received data \cite{quic-draft-11}.

\begin{figure*}
   \vspace{-0.35cm}
   \hspace{-0.8cm}
   \begin{minipage}[t]{.34\linewidth}
   \captionsetup{width=.95\linewidth}
\includegraphics[scale=0.19]{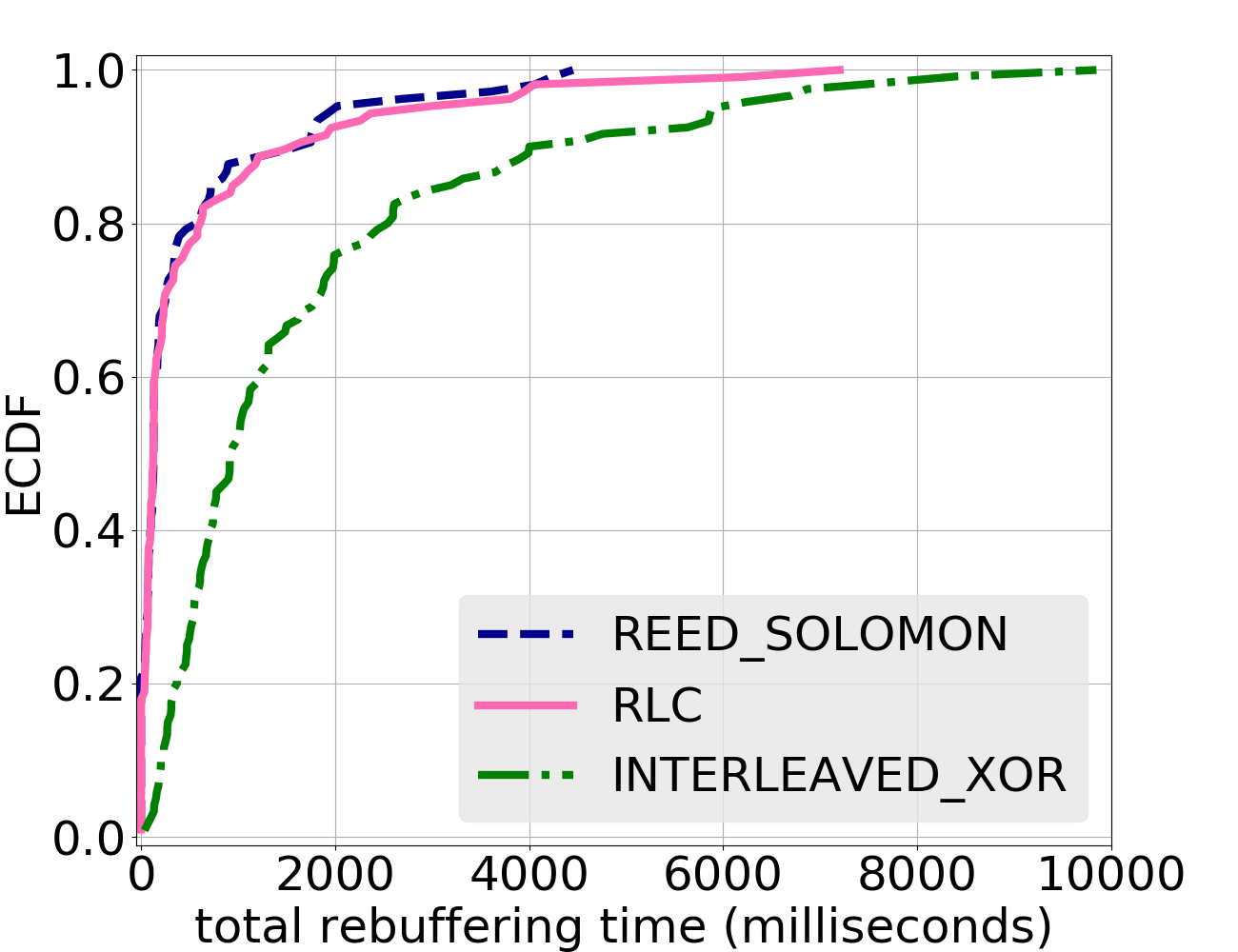}
   \caption{Reed-Solomon VS RLC VS interleaved XOR: total re-buffering time.}
   \label{fig_rs_rlc_xor_rebuffering}
   \end{minipage}
   \begin{minipage}[t]{.34\linewidth}
   \captionsetup{width=.95\linewidth}
   \includegraphics[scale=0.19]{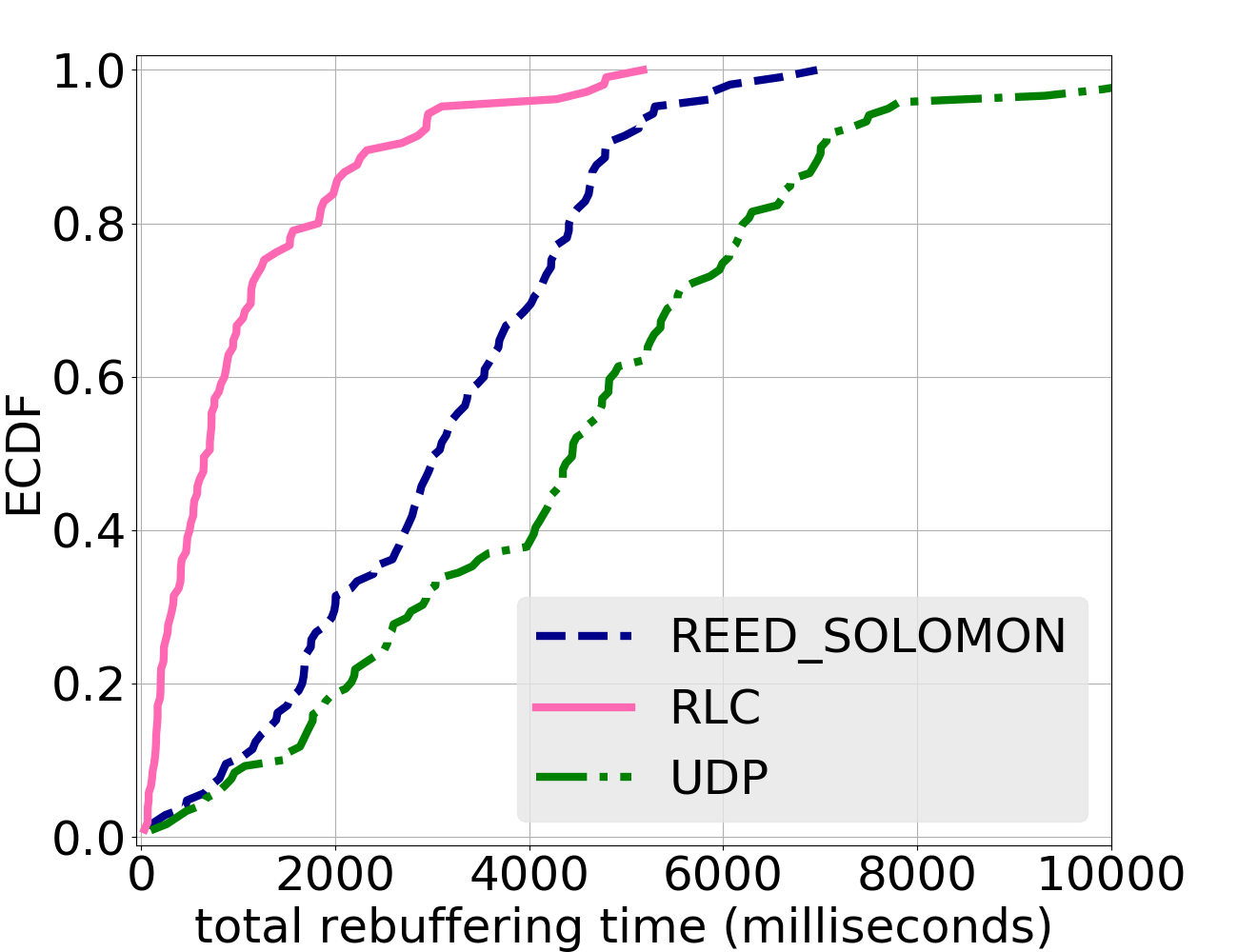}
   \caption{Reed-Solomon VS RLC VS UDP: total rebuffering time with low desynchronisation.}
   \label{fig_rs_rlc_xor_rebuffering_33ms}
   \end{minipage}
   \begin{minipage}[t]{.34\linewidth}
   \captionsetup{width=.95\linewidth}
   \includegraphics[scale=0.19]{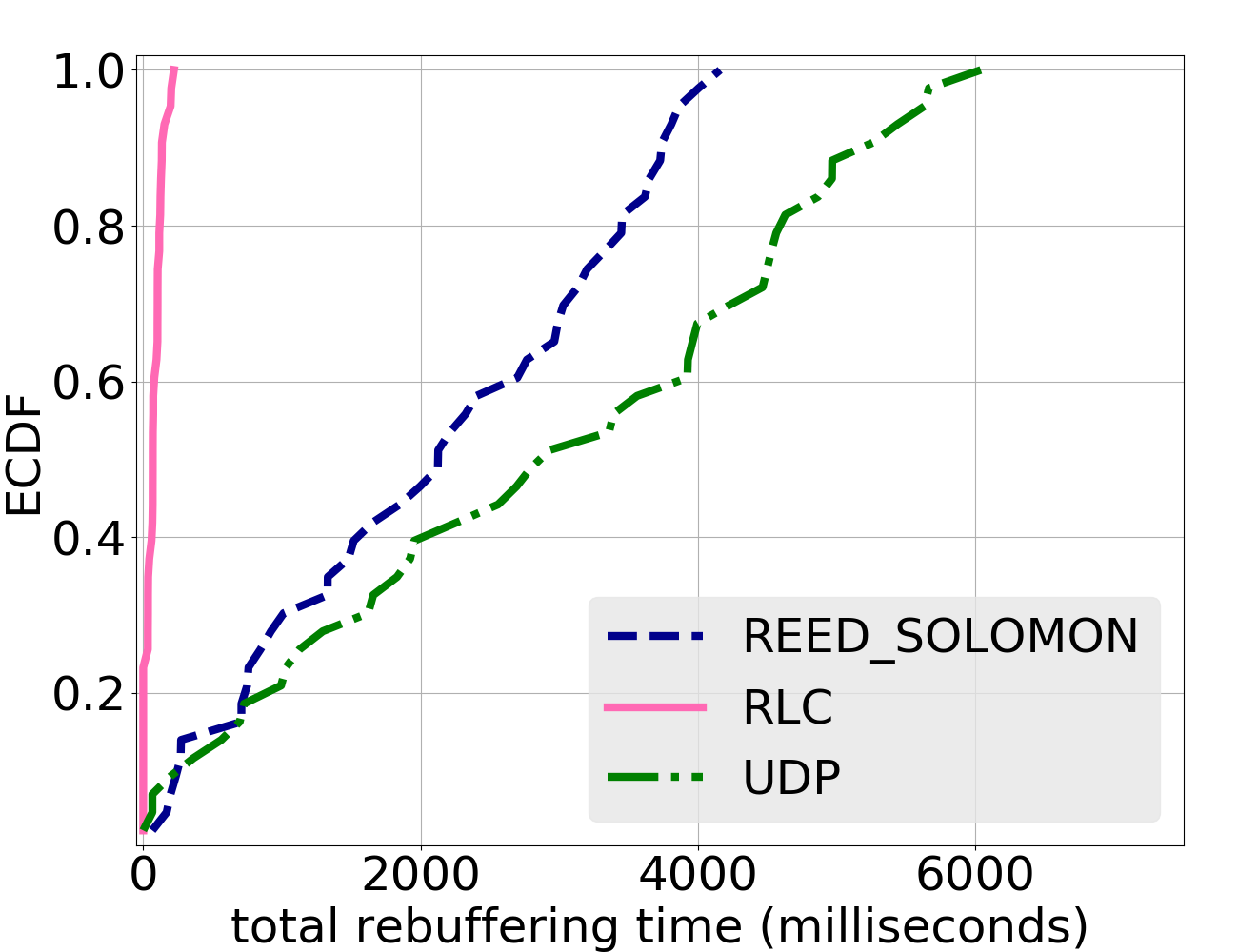}
   \caption{Reed-Solomon VS RLC VS UDP: total rebuffering time with low desynchronisation and uniform losses ranging from 0 to 3\%.}
   \label{fig_rs_rlc_xor_rebuffering_33ms_uniform}
   \end{minipage}
\end{figure*}


\subsection{Exploring FEC schemes}

In this section, we compare our different FEC Schemes: a $(30, 20)$ Reed-Solomon, a $(3, 2, 20)$ convolutional RLC and a $(3, 2)$ XOR FEC Scheme with 10 interleaved Source Blocks. Figure \ref{fig_rs_rlc_xor_rebuffering} shows the total re-buffering time using the three solutions. While Reed-Solomon and RLC seem to behave similarly, the XOR FEC Scheme recovers a lot less data. This is because even when using interleaving, the XOR FEC Scheme has poorer recovery capabilities. When looking more closely at our results, the Reed-Solomon FEC Scheme seems to encounter fewer rebufferings in the majority of the cases compared to RLC, especially with a long mean burst loss length. This is a drawback of convolutional FEC Schemes, they need a longer latency to recover from long bursts than block FEC Schemes. The convolutional RLC FEC Scheme however seems to encounter fewer rebufferings when the mean loss burst is shorter.

\subsubsection{Impact of the playback buffer}
In order to see the impact of the playback buffer on convolutional and block FEC Schemes, we performed tests with a playback buffer of 33 milliseconds instead of the default 100 milliseconds. We use Reed-Solomon to represent the block FEC Schemes as it provides better recovery capabilities than the XOR FEC Scheme. We use the convolutional RLC FEC Scheme to represent the convolutional FEC Schemes. We also analyse the performances of UDP in each experiment to see if the performances of QUIC-FEC come close to those of UDP because of the small playback buffer. Figure \ref{fig_rs_rlc_xor_rebuffering_33ms} shows the total re-buffering time with a desynchronisation of 33 milliseconds. In that configuration, the RLC FEC Scheme encounters fewer rebufferings than the Reed-Solomon FEC Scheme. This is due to the fact that with convolutional FEC Schemes, the Repair Symbols are interleaved with the Source Symbols while in our $(30, 20)$ block FEC Scheme, the Repair Symbols will be received after receiving the 20 Source Symbols.  When the loss bursts are shorter, it  needs less time to recover the lost packets. With a block FEC Scheme, it always takes the same amount of time to recover losses if they start at the same symbol: the FEC Scheme needs to wait for the end of the block to receive the Repair Symbols and be able to recover the packets. Reducing the de-synchronization will thus have a lower impact on convolutional FEC Schemes: the short loss bursts will still be recovered on time, while this will not necessarily be the case anymore for block FEC Schemes.

In order to better see the difference between convolutional and block FEC Schemes, we performed experiments with uniformly distributed losses. We use a loss rate in the range $[0, 0.03]$, in line with the range proposed by Paasch \textit{et al.} \cite{experimentaldesignmptcp}. As there are now two dimensions in the parameter space when using the uniform loss model, we run the experiments with 40 different combinations of parameters instead of 120. It reduces the time needed to perform our experiments while subsampling sufficiently the parameter space. Figure \ref{fig_rs_rlc_xor_rebuffering_33ms_uniform} shows the total re-buffering time when using the uniform loss model.
We can see that the RLC FEC Scheme encounters fewer re-bufferings in the vast majority of the cases. This is an advantage of convolutional FEC Schemes: they are able to nearly instantly recover from isolated losses. On the other hand, the block FEC Schemes always have to wait for the reception of the end of the block to be able to recover lost packets.

\section{Multipath Experiments}

\begin{figure}
\includegraphics[scale=0.6]{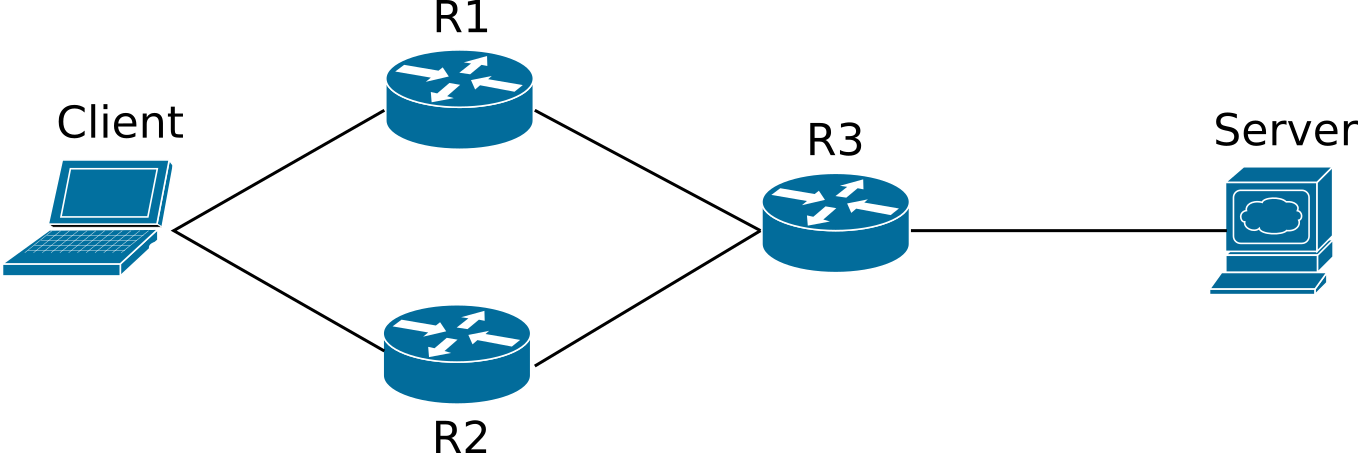}
\caption{Multipath topology used in the experiments.}
\label{fig_mp_config}
\end{figure}

In this section, we perform experiments using the Multipath QUIC implementation in \texttt{quic-go} \cite{mpquic} to explore the possibilities brought by a multipath transfer when using Forward Erasure Correction. More specifically, we explore an interesting property linking block FEC Schemes and multipath interleaving. Figure \ref{fig_mp_config} represents the multipath topology used for the experiments. This configuration offers two network interfaces to the client and only one for the server. This results in two different possible paths, the path \textit{Client-R1-R3-Server} that we call \textit{path$_1$} and the path \textit{Client-R2-R3-Server} that we call \textit{path$_2$}. The single-path client always contacts the server using \textit{path$_1$}.
Unless stated otherwise, the FEC Scheme used in the experiments is the $(30, 20)$ Reed-Solomon block FEC Scheme of the previous sections.

In this section, since there are twice more dimensions in the parameter space, all the experiments are performed with a simplified Gilbert-Elliott model with $h = 0$ and $k = 1$.
The remaining parameters are still randomly selected in the ranges described in Table \ref{table_parameters_range}. We consider similar delays for both paths. The loss models are applied independently on links $\operatorname{R1-R3}$ and $\operatorname{R2-R3}$.

\begin{figure*}[t]
	\vspace{-0.4cm}
	\hspace{-0.8cm}
	\begin{minipage}[t]{.34\linewidth}
		\captionsetup{width=.95\linewidth}
		\includegraphics[scale=0.19]{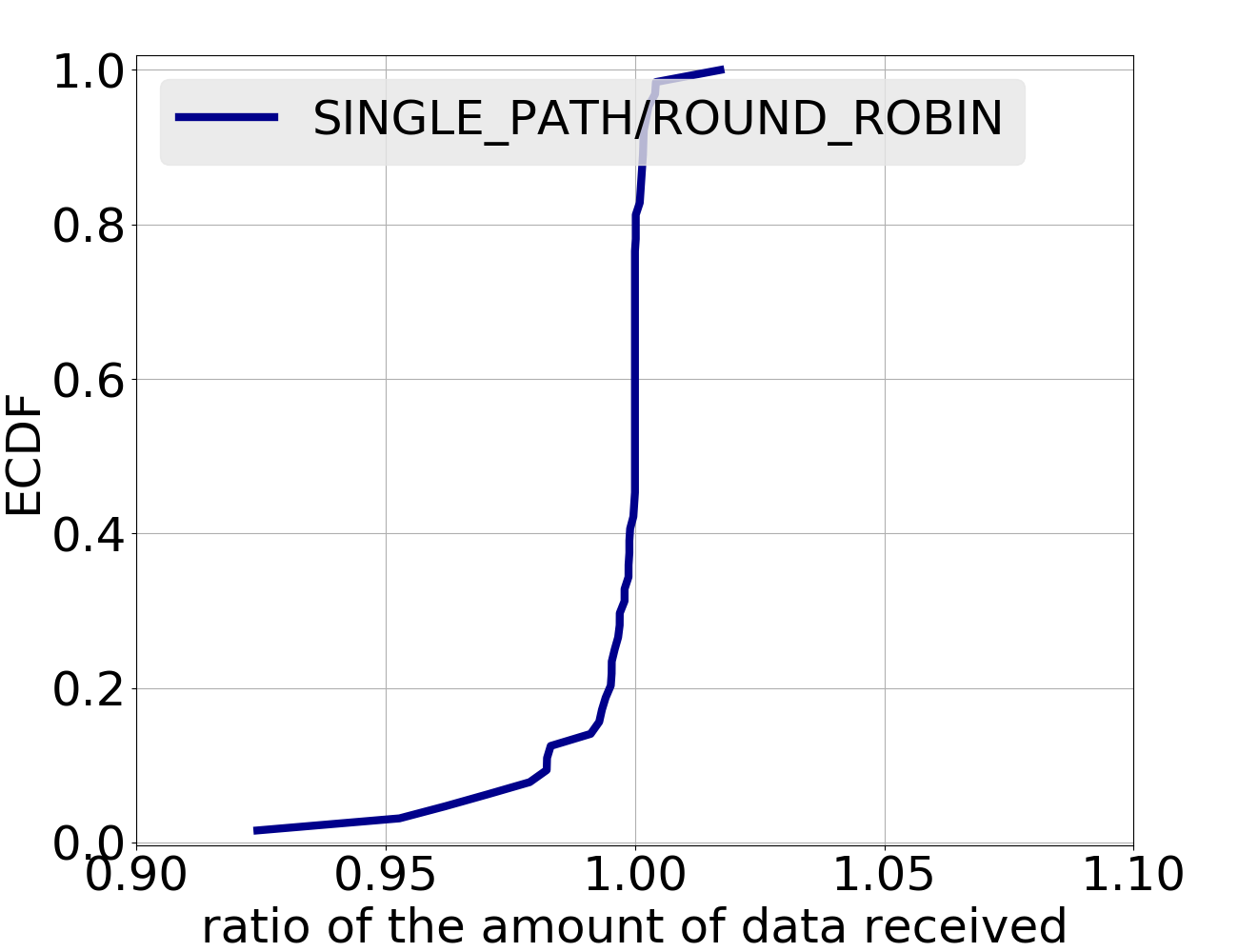}
		\caption{Round-robin VS single-path with Reed-Solomon: ratio of the amount of application data received.}
		\label{fig_roundrobin_singlepath_ratio_homogeneous}
	\end{minipage}
	\begin{minipage}[t]{.34\linewidth}
		\captionsetup{width=.95\linewidth}
		\includegraphics[scale=0.19]{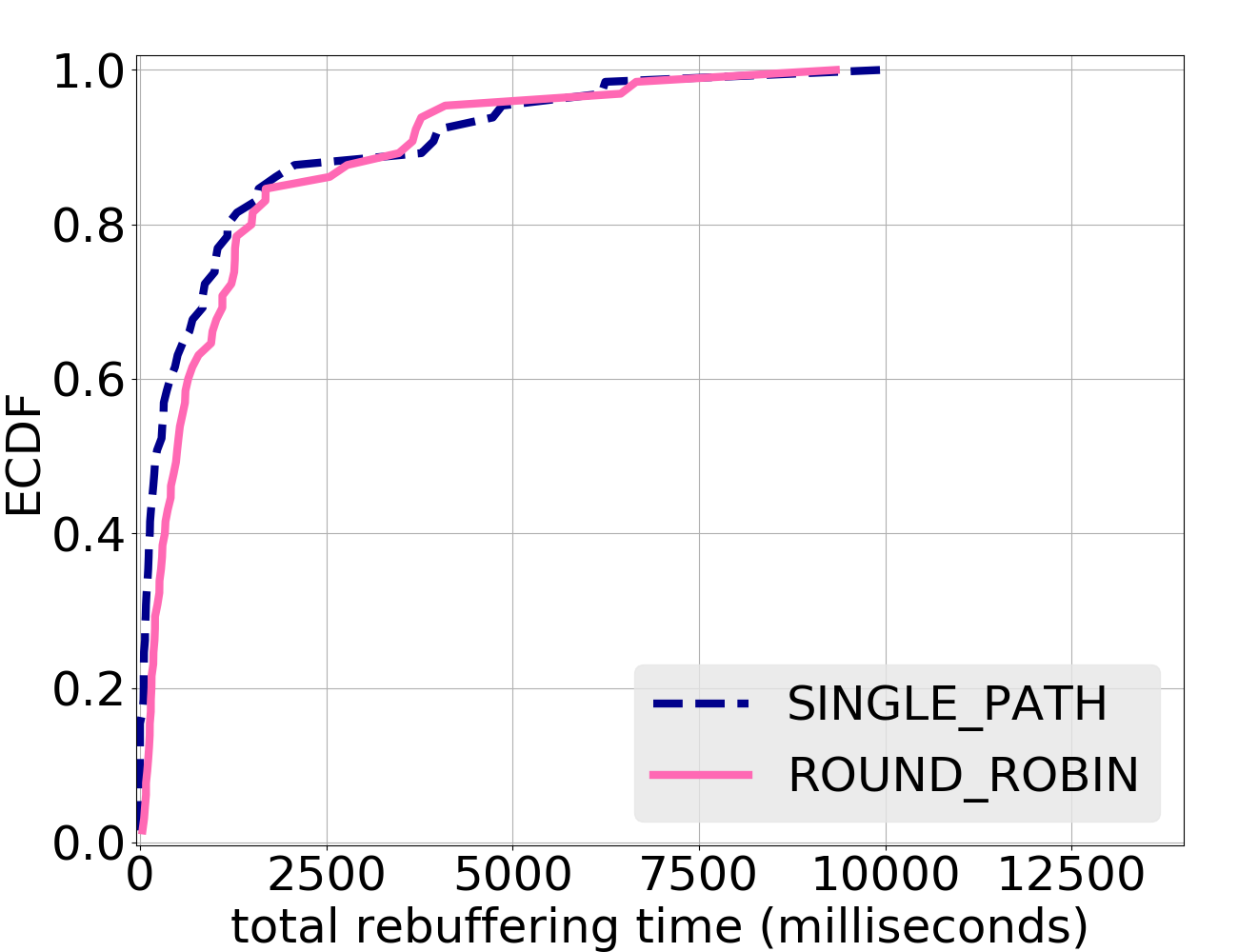}
		\caption{Round-robin VS single-path VS HIGHRB scheduler: total rebuffering time with homogeneous paths.}
		\label{fig_roundrobin_singlepath_rebuffering_homogeneous}
	\end{minipage}
	\begin{minipage}[t]{.34\linewidth}
		\captionsetup{width=.95\linewidth}
		\includegraphics[scale=0.19]{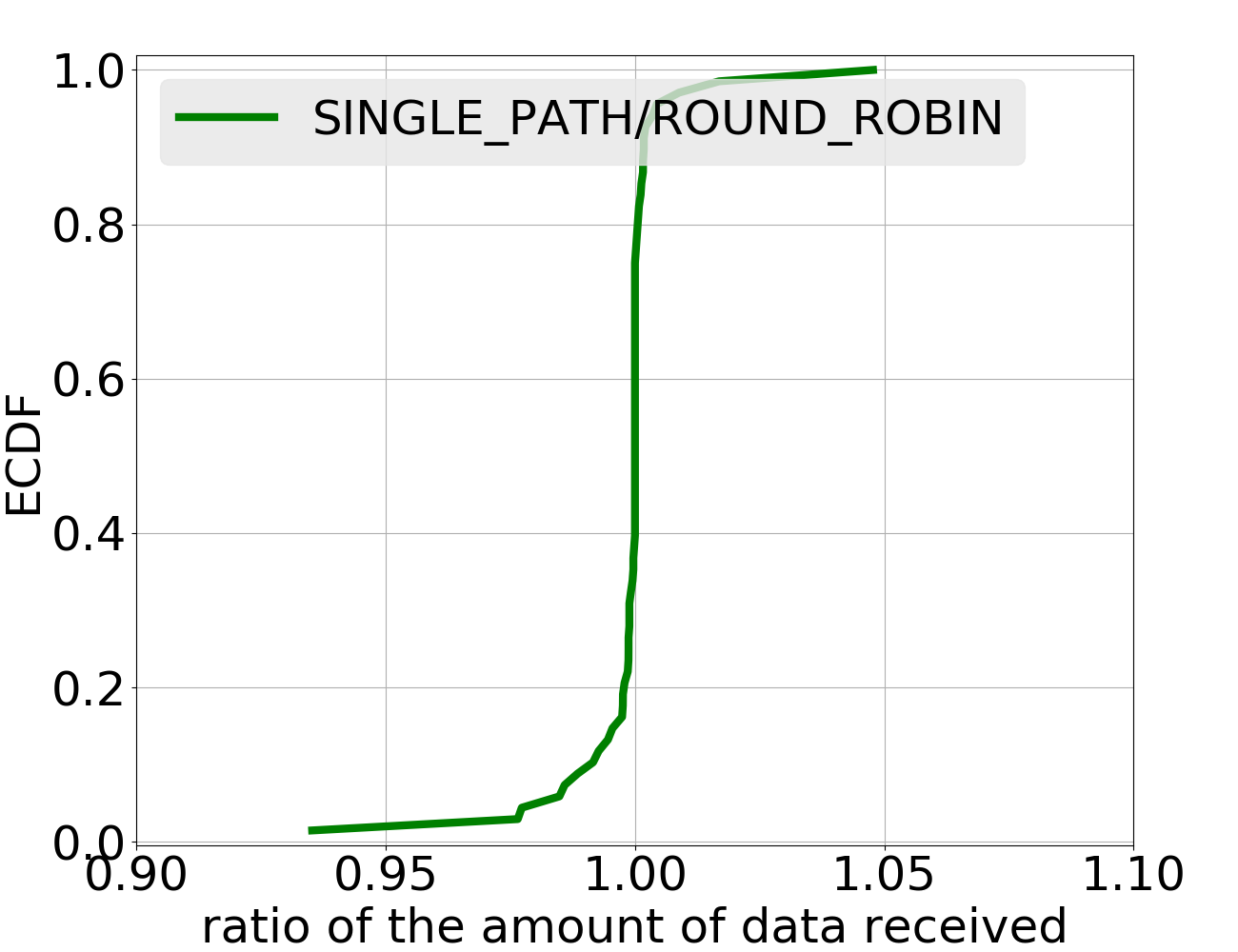}
		\caption{Round-robin VS single-path scheduler with RLC: ratio of the amount of application data received.}
		\label{fig_rlc_roundrobin_singlepath_ratio_homogeneous}
	\end{minipage}
\end{figure*}

\subsection{The Potential Benefits of Multipath with FEC}

\begin{figure}[t]
\centering
\includegraphics[scale=0.45]{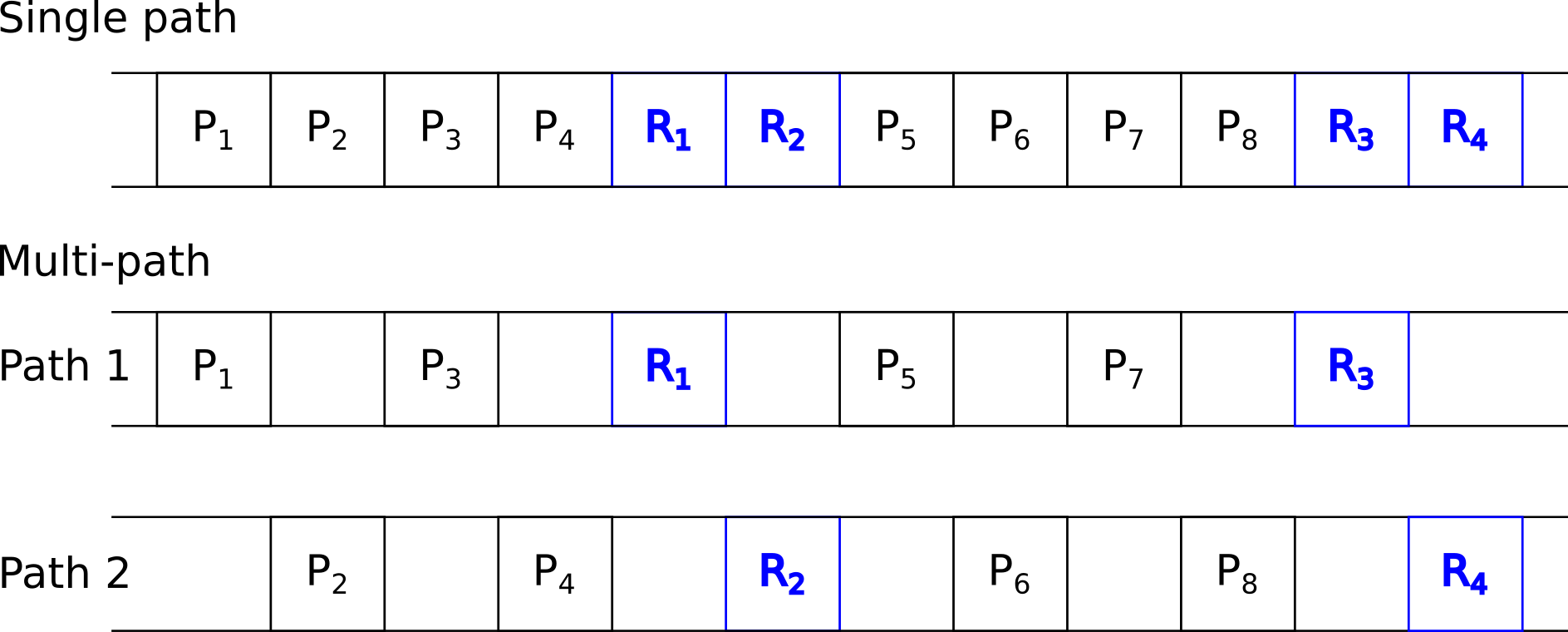}
\caption{Bursts with single-path VS multipath with a round-robin scheduler}
\label{fig_mp_bursts}
\end{figure}

When considering burst losses, sending the data on different paths can help to recover more packets with the same level of redundancy compared to single-path transmission. An example is shown in Figure \ref{fig_mp_bursts}. In a single-path configuration, a burst of 3 symbols will be recovered if the burst begins at $R_1$, $R_2$, $R_3$ or $R_4$ and will be lost in every other cases (if the burst begins at any of the 8 other symbols). It will thus be recovered in $\frac{1}{3}$ of the cases. With a multipath configuration using a round-robin scheduler and considering that losses occurring on both paths at the same time are rare, a burst loss of 3 symbols on Path 1 will be recovered if the burst begins at $P_3$, $R_1$, $P_7$ or $R_3$ and will be lost if the burst begins at $P_1$ or $P_5$. It will thus be recovered in 2/3 of the cases. Indeed, when a burst loss affects on one path the packets of one Source Block, some packets of the Source Block are still received as they are sent on the other path. Interleaving the packets thus spreads the losses on multiple Source Blocks that are handled independently.

\subsubsection{Comparing round-robin with single-path with homogeneous paths}
We first perform experiments using homogeneous paths to verify these properties. As the paths use a simplified Gilbert-Elliott model and are homogeneous, we run the experiments with 60 different combinations of parameters. Figure \ref{fig_roundrobin_singlepath_ratio_homogeneous} directly compares the amount of data received with the round-robin scheduler and the single-path case. 
As expected, the total amount of received data is often higher in the multipath case. When looking at our results, we saw that for all the cases where single-path performed better with a ratio above 1.001, it was either with parameter configuration with a high value for the $r$ parameter ($\geq 0.3$) or a low value for both the $r$ and $p$ parameters, making both approaches having quite similar results. On the other hand, we can see in Figure \ref{fig_roundrobin_singlepath_rebuffering_homogeneous} that the total re-buffering time can be longer with multipath compared to single-path. By looking at our results, the multipath scenario has a smaller total re-buffering time when the mean burst loss is short and a longer total re-buffering time when the mean loss burst is long. This is the drawback of interleaving the packets on the paths. When interleaving the packets of two paths, a loss of $n$ packets will be spread on more application messages than in the single-path case. If these bursts are not recovered, this will imply the corruption of more application messages and thus more rebufferings.

The benefits of multipath are thus not completely clear in our use-case in terms of the total re-buffering time. However, the amount of data received is still an interesting metric as it can be important for applications whose messages are not longer than one packet. 

We also analyse the benefits of performing interleaving with our convolutional RLC FEC Scheme. Figure \ref{fig_rlc_roundrobin_singlepath_ratio_homogeneous} shows the ratio of the amount of data received between the single-path RLC FEC Scheme and the same FEC Scheme used with a round-robin scheduler. We can see that compared to the Reed-Solomon FEC Scheme, the advantage of multipath is less visible. Indeed, interleaving the packets during a loss burst virtually increases the length of the burst. While it spreads the burst in several \textit{independently} decoded blocks for block FEC Scheme, it implies a longer burst among inter-dependent equations in the case of the RLC FEC Scheme. On the other hand, it needs to receive fewer Repair Symbols after the end of the burst, as Source Symbols are received during the loss burst. This mitigates the drawback of spreading the burst on a longer period. 
Based on this result, we will thus focus on the Reed-Solomon FEC Scheme in the remaining experiments.

\subsection{The Need for an Adapted Scheduler}

The previous section compared the round-robin scheduler to the single-path case with homogeneous paths. Figure \ref{fig_roundrobin_highrb_singlepath_heterogeneous} compares the amount of application data received using paths with potentially different values for the parameters of their loss models. As the paths are now heterogeneous, adding new dimensions to the parameter space, we now run the experiments with 120 different combinations of parameters. In that case, the advantage of the multipath interleaving with round-robin is not clear anymore compared to single-path. By looking at our results, we see that for the cases where the round-robin scheduler received more data, the $p$ parameter of \textit{path$_1$} is greater than the $p$ parameter of \textit{path$_2$} by $0.0018$ and the $r$ parameter is lower by $0.19$ on average. This implies more packet losses and longer loss bursts, leading to more unrecoverable losses on average. This explains why the round-robin scheduler performs better in these cases. When \textit{path$_2$} is better than \textit{path$_1$} in terms of losses, using both paths is better than only using the worse. We also have the same phenomenon for the cases where the single-path scenario received more data: in these cases, the $p$ parameter of \textit{path$_1$} is lower by $0.0019$ and $r$ is greater by $0.15$ on average compared to \textit{path$_2$}, meaning that only using \textit{path$_1$} is better than using both paths. This means that we need an adaptive scheduler, using the best path when the paths are different and performing interleaving when the paths are similar.

\subsection{The HighRB Scheduler}

In this section, we propose a new scheduler whose purpose is to perform path interleaving when possible, and use only one path when using both could be harmful. We define the HighRB scheduler as a scheduler that selects a path randomly, using the number of remaining bytes of the congestion window computed by a congestion control scheme as weights of the random selection. The scheduler is described in Algorithm \ref{algo_highrb}.

\begin{algorithm}
\caption{HighRB scheduler}\label{algo_highrb}
\begin{algorithmic}[1]
\Require $ \mathcal P $, the set of available paths
\Require $BytesInFlight(p)$, the number of sent, non-acknowledged bytes of $p \, , \, \forall p \in \mathcal{P}$
\Require $Cwin(p)$, the congestion window of $p\, , \, \forall p \in \mathcal{P}$
\State $RB(p) \gets Cwin(p) - BytesInFlight(p)\, , \,  \forall p \in \mathcal P$
\State $Total \gets \sum_{p \in \mathcal{P}} RB(p)$
\If {$Total \neq 0$}
    \State $w(p) \gets \frac{RB(p)}{Total}\, , \, \forall p \in \mathcal P$
\Else
    \State $w(p) \gets \frac{1}{|\mathcal{P}|}\, , \, \forall p \in \mathcal P$
\EndIf
\State select $p^*$ randomly from $\mathcal{P}$ with weights $w$
\State \Return $p^*$
\end{algorithmic}
\end{algorithm}

The purpose of this scheduler is to gather information from a loss-based congestion control algorithm. The congestion window of lossy paths will be smaller than the congestion window of other paths, implying a lower number of remaining bytes on average and a lower probability to select this path. Our implementation of HighRB in \texttt{quic-go} relies on the congestion window provided by the OLIA congestion control algorithm \cite{olia}. HighRB uses a weighted random selection instead of always selecting the path with the highest number of remaining bytes since we focus on the real-time use-case and real-time applications do not often use all the available bandwidth. Selecting always the path with the highest number of remaining bytes could thus lead to use only one path instead of doing interleaving if the sending rate of the application is not large enough. Finally, the random selection also allows continuing to regularly probe the non-preferred paths to check if it could become an interesting path to use.


\begin{figure}[t]
	\captionsetup{width=.\linewidth}
	\includegraphics[scale=0.23]{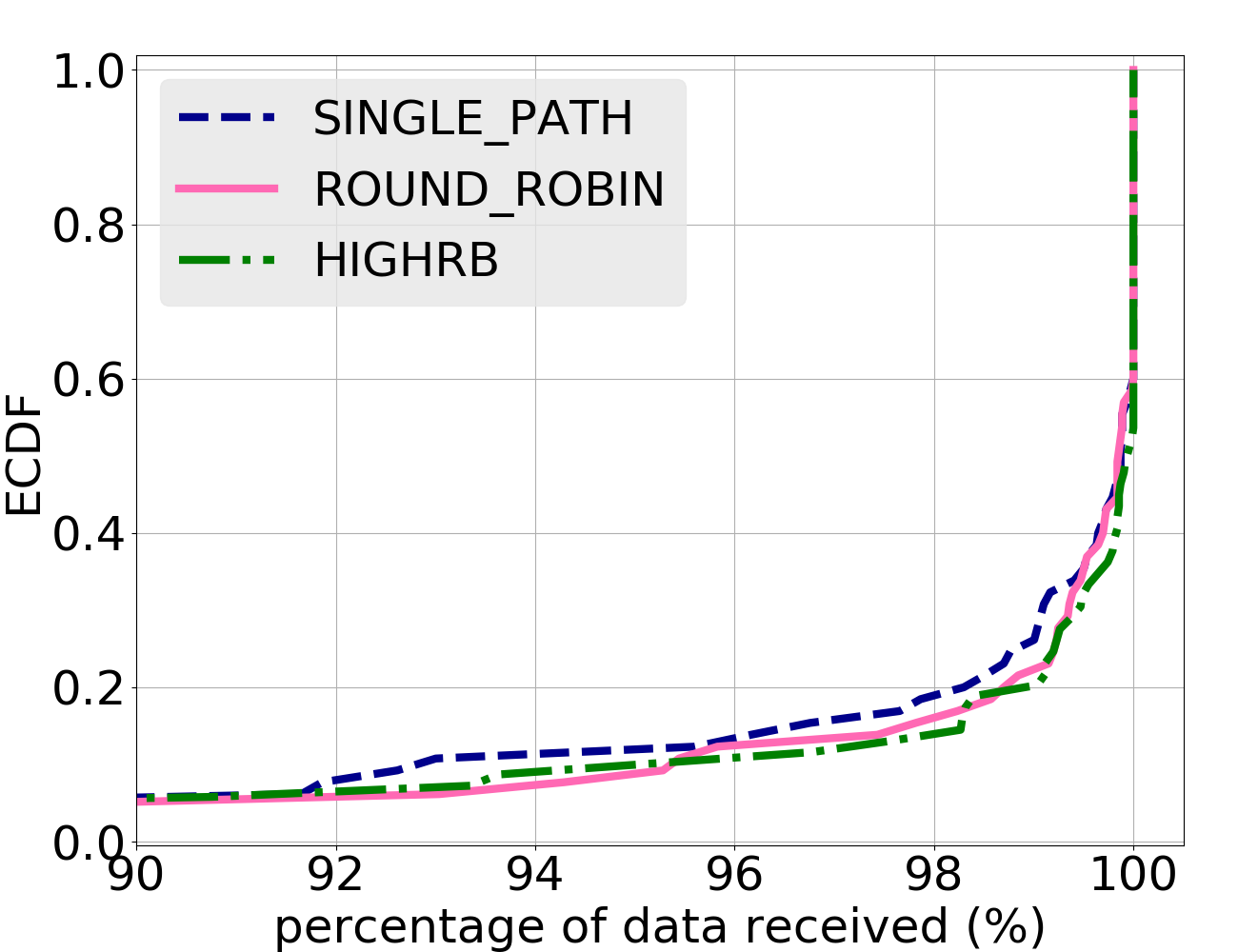}
	\caption{Round-robin VS single-path VS HighRB scheduler: amount of application data received with loss-homogeneous paths.}
	\label{fig_roundrobin_highrb_singlepath_homogeneous}
\end{figure}

\subsubsection{Performances of HighRB}

Figure \ref{fig_roundrobin_highrb_singlepath_homogeneous} shows 
the performance of HighRB when the paths are homogeneous.  We can see that HighRB seems to perform quite similarly than the round-robin scheduler. When looking more closely at our data and comparing HighRB and the round-robin scheduler, HighRB however received more data in every configuration with a low delay ($\leq 15ms$). 
This is an advantage of using the remaining bytes of the congestion window as a metric to choose a path. As we send one application message every 33 milliseconds, a long loss burst that spans two application messages will be taken into account by the congestion control before its end when the round-trip-time is short. It will thus lower the congestion window and the number of remaining bytes before the end of the burst. The scheduler will choose the other path on average to avoid to loose additional packets, while the round-robin scheduler will continue to use both paths.

\begin{figure}[t]
	\captionsetup{width=\linewidth}
	\includegraphics[scale=0.23]{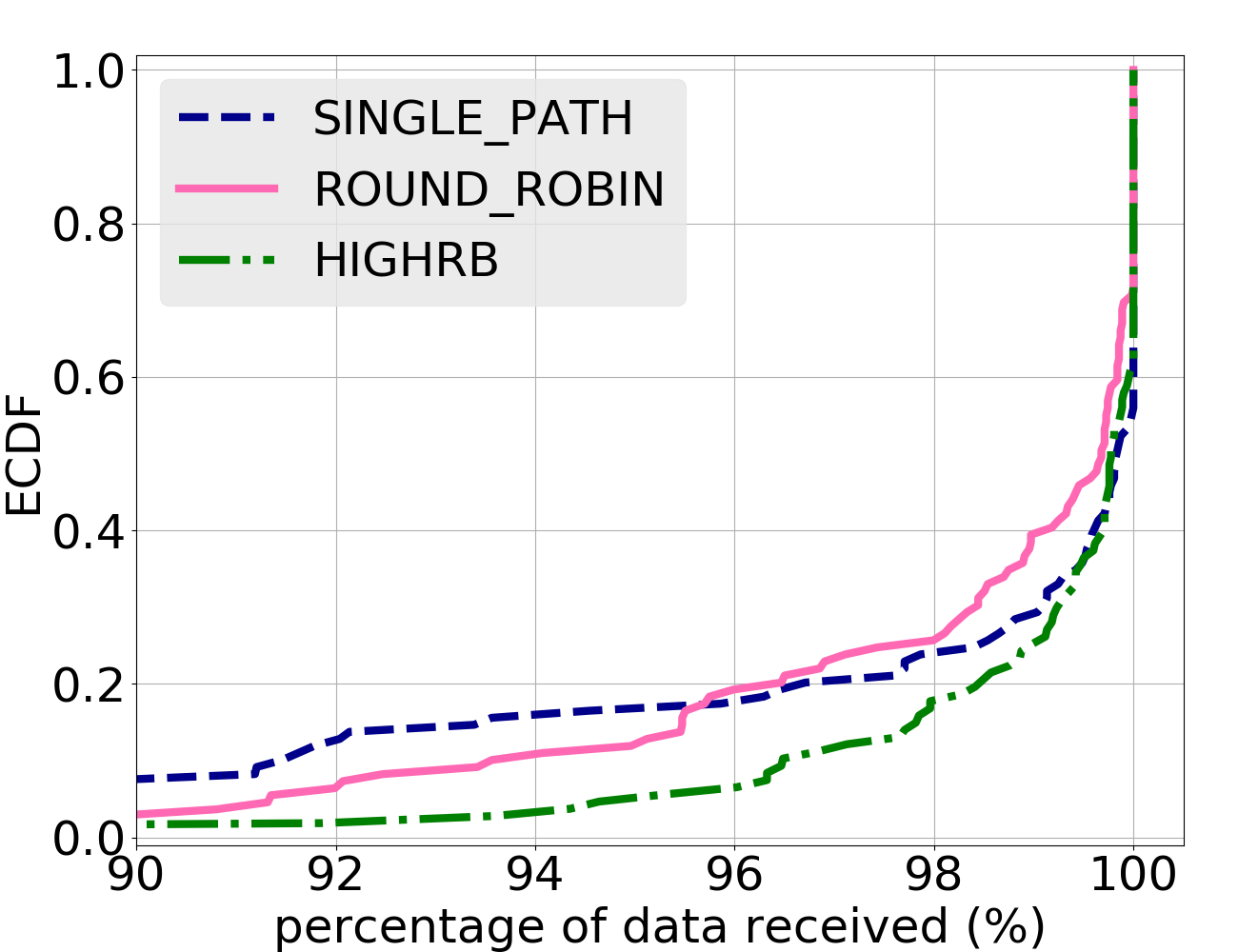}
	\caption{Round-robin VS single-path VS HighRB scheduler: amount of application data received with loss-heterogeneous paths.}
	\label{fig_roundrobin_highrb_singlepath_heterogeneous}
\end{figure}

Finally, we take a look at the performances of HighRB when the paths can have different loss parameters. Figure \ref{fig_roundrobin_highrb_singlepath_heterogeneous} shows the ECDF of the amount of data received compared to single-path and the round-robin scheduler. We can see that HighRB seems to outperform both other alternatives. We found however some cases where the single-path solution performed sensibly better than HighRB. These cases were cases for which the $r$ parameter of \textit{path$_1$} was strongly higher than for \textit{path$_2$}. In that case, although HighRB is adaptive, only using the first path gives better results.

%% file: conclusion.tex
\section{Conclusion}


While QUIC was initially designed for the HTTP/2 use case, we believe that other applications can benefit from this new transport protocol.
One of such applications is traffic with strict deadlines, such as live video streaming.
In this paper, we presented QUIC-FEC as an extension of the current QUIC protocol.
It provides unreliable service delivery and connection-defined Forward Erasure Correction while keeping desirable QUIC features like encryption and protocol extensibility.
We implemented QUIC-FEC on top of \texttt{quic-go} and used it to evaluate three FEC schemes.
Through extensive evaluation with emulations covering a wide range of networks parameters with bursty and uniform losses, we demonstrated the benefits of applying FEC to unreliable traffic.
Our results indicate that block and convolutional based FEC schemes are more adapted than XOR-based ones in such configurations.
We also explored how the usage of multiple paths can help in increasing the delivery rate without increasing the generated traffic, and proposed a packet scheduler aiming to achieve this goal.
Although we observed that the delivery delay can increases, we showed that spreading symbols over several paths can effectively provide higher data delivery rates.

We believe our work will benefit to both transport protocol and FEC communities.
Our future work includes exploration of other FEC schemes and in-the-wild experiments.


\section*{Artefacts}

To ensure the reproducibility of our work, we will make the sources of our implementation available publicly.
Currently, these are accessible by contacting the authors.

%% file: paper.bbl

\begin{thebibliography}{55}


\ifx \showCODEN    \undefined \def \showCODEN     #1{\unskip}     \fi
\ifx \showDOI      \undefined \def \showDOI       #1{#1}\fi
\ifx \showISBNx    \undefined \def \showISBNx     #1{\unskip}     \fi
\ifx \showISBNxiii \undefined \def \showISBNxiii  #1{\unskip}     \fi
\ifx \showISSN     \undefined \def \showISSN      #1{\unskip}     \fi
\ifx \showLCCN     \undefined \def \showLCCN      #1{\unskip}     \fi
\ifx \shownote     \undefined \def \shownote      #1{#1}          \fi
\ifx \showarticletitle \undefined \def \showarticletitle #1{#1}   \fi
\ifx \showURL      \undefined \def \showURL       {\relax}        \fi
\providecommand\bibfield[2]{#2}
\providecommand\bibinfo[2]{#2}
\providecommand\natexlab[1]{#1}
\providecommand\showeprint[2][]{arXiv:#2}

\bibitem[\protect\citeauthoryear{??}{net}{[n. d.]}]%
        {netem}
 \bibinfo{year}{[n. d.]}\natexlab{}.
\newblock \bibinfo{title}{netem}.
\newblock
  \bibinfo{howpublished}{\url{https://wiki.linuxfoundation.org/networking/netem}}.
    (\bibinfo{year}{[n. d.]}).
\newblock
\newblock
\shownote{Accessed: 2018-05-31.}


\bibitem[\protect\citeauthoryear{??}{qui}{[n. d.]a}]%
        {quicfecprague}
 \bibinfo{year}{[n. d.]}\natexlab{a}.
\newblock \bibinfo{title}{QUIC-FEC}.
\newblock
  \bibinfo{howpublished}{\url{https://datatracker.ietf.org/meeting/99/materials/slides-99-nwcrg-08-swett-quic-fec-00}}.
    (\bibinfo{year}{[n. d.]}).
\newblock
\newblock
\shownote{Accessed: 2018-06-02.}


\bibitem[\protect\citeauthoryear{??}{qui}{[n. d.]b}]%
        {quicdtls}
 \bibinfo{year}{[n. d.]}\natexlab{b}.
\newblock \bibinfo{title}{QUIC over DTLS}.
\newblock
  \bibinfo{howpublished}{\url{https://github.com/quicwg/base-drafts/issues/1165}}.
    (\bibinfo{year}{[n. d.]}).
\newblock
\newblock
\shownote{Accessed: 2018-06-20.}


\bibitem[\protect\citeauthoryear{??}{gqu}{[n. d.]}]%
        {gquic}
 \bibinfo{year}{[n. d.]}\natexlab{}.
\newblock \bibinfo{title}{QUIC Wire Layout Specification}.
\newblock
  \bibinfo{howpublished}{\url{https://docs.google.com/document/d/1WJvyZflAO2pq77yOLbp9NsGjC1CHetAXV8I0fQe-B_U/edit}}.
    (\bibinfo{year}{[n. d.]}).
\newblock
\newblock
\shownote{Accessed: 2018-04-24.}


\bibitem[\protect\citeauthoryear{A.Antonovs}{A.Antonovs}{2015}]%
        {Antonov_go-math}
\bibfield{author}{\bibinfo{person}{A.Antonovs}.}
  \bibinfo{year}{2015}\natexlab{}.
\newblock \bibinfo{title}{Extra mathematical algorithms and data types for
  Golang}.  (\bibinfo{year}{2015}).
\newblock
\newblock
\shownote{\url{https://github.com/alex-ant/gomath}.}


\bibitem[\protect\citeauthoryear{Adamson et~al\mbox{.}}{Adamson
  et~al\mbox{.}}{2018}]%
        {Adamson_Taxonomy:2018}
\bibfield{author}{\bibinfo{person}{B. Adamson} {et~al\mbox{.}}}
  \bibinfo{year}{2018}\natexlab{}.
\newblock \bibinfo{title}{Taxonomy of Coding Techniques for Efficient Network
  Communications}.  (\bibinfo{date}{March} \bibinfo{year}{2018}).
\newblock
\newblock
\shownote{Internet draft, draft-irtf-nwcrg-network-coding-taxonomy-08, work in
  progress.}


\bibitem[\protect\citeauthoryear{Badr, Khisti, Tan, and Apostolopoulos}{Badr
  et~al\mbox{.}}{2013}]%
        {badr2013streaming}
\bibfield{author}{\bibinfo{person}{Ahmed Badr}, \bibinfo{person}{Ashish
  Khisti}, \bibinfo{person}{Wai-Tian Tan}, {and} \bibinfo{person}{John
  Apostolopoulos}.} \bibinfo{year}{2013}\natexlab{}.
\newblock \showarticletitle{Streaming codes for channels with burst and
  isolated erasures}. In \bibinfo{booktitle}{{\em INFOCOM, 2013 Proceedings
  IEEE}}. IEEE, \bibinfo{pages}{2850--2858}.
\newblock


\bibitem[\protect\citeauthoryear{Belshe and Peon}{Belshe and Peon}{[n. d.]}]%
        {spdy}
\bibfield{author}{\bibinfo{person}{M. Belshe} {and} \bibinfo{person}{R. Peon}.}
  \bibinfo{year}{[n. d.]}\natexlab{}.
\newblock \bibinfo{title}{SPDY protocol}.
\newblock
  \bibinfo{howpublished}{\url{http://www.chromium.org/spdy/spdy-protocol/spdy-protocol-draft3-1}}.
    (\bibinfo{year}{[n. d.]}).
\newblock
\newblock
\shownote{Accessed: 2018-06-19.}


\bibitem[\protect\citeauthoryear{Biersack}{Biersack}{1993}]%
        {biersack1993performance}
\bibfield{author}{\bibinfo{person}{Ernst~W Biersack}.}
  \bibinfo{year}{1993}\natexlab{}.
\newblock \showarticletitle{Performance evaluation of forward error correction
  in an ATM environment}.
\newblock \bibinfo{journal}{{\em IEEE Journal on Selected Areas in
  Communications\/}} \bibinfo{volume}{11}, \bibinfo{number}{4}
  (\bibinfo{year}{1993}), \bibinfo{pages}{631--640}.
\newblock


\bibitem[\protect\citeauthoryear{Bishop}{Bishop}{2018}]%
        {quic-http}
\bibfield{author}{\bibinfo{person}{Mike Bishop}.}
  \bibinfo{year}{2018}\natexlab{}.
\newblock \bibinfo{booktitle}{{\em Hypertext Transfer Protocol (HTTP) over
  QUIC}}.
\newblock \bibinfo{type}{Internet-Draft} draft-ietf-quic-http-11.
  \bibinfo{institution}{IETF Secretariat}.
\newblock
\showURL{%
\url{http://www.ietf.org/internet-drafts/draft-ietf-quic-http-11.txt}}


\bibitem[\protect\citeauthoryear{Bob~Lantz and McKeown}{Bob~Lantz and
  McKeown}{2010}]%
        {mininet}
\bibfield{author}{\bibinfo{person}{Brandon~Heller Bob~Lantz} {and}
  \bibinfo{person}{Nick McKeown}.} \bibinfo{year}{2010}\natexlab{}.
\newblock \showarticletitle{A network in a laptop: rapid prototyping for
  software-defined networks}. In \bibinfo{booktitle}{{\em Proceedings of the
  9th ACM SIGCOMM Workshop on Hot Topics in Networks}}. ACM,
  \bibinfo{pages}{19}.
\newblock


\bibitem[\protect\citeauthoryear{Bouabdallah, Cunche, Roca, Matsuzono, and
  Lacan}{Bouabdallah et~al\mbox{.}}{2013}]%
        {reedsolomonfecframe}
\bibfield{author}{\bibinfo{person}{Amine Bouabdallah}, \bibinfo{person}{Mathieu
  Cunche}, \bibinfo{person}{Vincent Roca}, \bibinfo{person}{Kazuhisa
  Matsuzono}, {and} \bibinfo{person}{Jerome Lacan}.}
  \bibinfo{year}{2013}\natexlab{}.
\newblock \showarticletitle{RFC6865: Simple Reed-Solomon Forward Error
  Correction (FEC) Scheme for FECFRAME}.
\newblock  (\bibinfo{year}{2013}).
\newblock


\bibitem[\protect\citeauthoryear{Carle and Biersack}{Carle and
  Biersack}{1997}]%
        {carle1997survey}
\bibfield{author}{\bibinfo{person}{Georg Carle} {and} \bibinfo{person}{Ernst~W
  Biersack}.} \bibinfo{year}{1997}\natexlab{}.
\newblock \showarticletitle{Survey of error recovery techniques for IP-based
  audio-visual multicast applications}.
\newblock \bibinfo{journal}{{\em IEEE Network\/}} \bibinfo{volume}{11},
  \bibinfo{number}{6} (\bibinfo{year}{1997}), \bibinfo{pages}{24--36}.
\newblock


\bibitem[\protect\citeauthoryear{Carlucci, De~Cicco, and Mascolo}{Carlucci
  et~al\mbox{.}}{2015}]%
        {quicexperiments}
\bibfield{author}{\bibinfo{person}{Gaetano Carlucci}, \bibinfo{person}{Luca
  De~Cicco}, {and} \bibinfo{person}{Saverio Mascolo}.}
  \bibinfo{year}{2015}\natexlab{}.
\newblock \showarticletitle{HTTP over UDP: an Experimental Investigation of
  QUIC}. In \bibinfo{booktitle}{{\em Proceedings of the 30th Annual ACM
  Symposium on Applied Computing}}. ACM, \bibinfo{pages}{609--614}.
\newblock


\bibitem[\protect\citeauthoryear{Cisco}{Cisco}{2017}]%
        {VNI:2017}
\bibfield{author}{\bibinfo{person}{Cisco}.} \bibinfo{year}{2017}\natexlab{}.
\newblock \bibinfo{title}{Cisco Visual Networking Index: Global Mobile Data
  Traffic Forecast Update, 2016–2021 r}.  (\bibinfo{date}{March}
  \bibinfo{year}{2017}).
\newblock
\newblock
\shownote{White Paper,
  \url{https://www.cisco.com/c/en/us/solutions/collateral/service-provider/visual-networking-index-vni/mobile-white-paper-c11-520862.html}.}


\bibitem[\protect\citeauthoryear{Coninck and Bonaventure}{Coninck and
  Bonaventure}{2018}]%
        {mpquic-design}
\bibfield{author}{\bibinfo{person}{Quentin~De Coninck} {and}
  \bibinfo{person}{Olivier Bonaventure}.} \bibinfo{year}{2018}\natexlab{}.
\newblock \bibinfo{booktitle}{{\em Multipath Extension for QUIC}}.
\newblock \bibinfo{type}{Internet-Draft} draft-deconinck-quic-multipath-00.
  \bibinfo{institution}{IETF Secretariat}.
\newblock
\showURL{%
\url{http://www.ietf.org/internet-drafts/draft-deconinck-quic-multipath-00.txt}}


\bibitem[\protect\citeauthoryear{Cui, Wang, Wang, Wang, and Wang}{Cui
  et~al\mbox{.}}{2015}]%
        {fmtcp}
\bibfield{author}{\bibinfo{person}{Yong Cui}, \bibinfo{person}{Lian Wang},
  \bibinfo{person}{Xin Wang}, \bibinfo{person}{Hongyi Wang}, {and}
  \bibinfo{person}{Yining Wang}.} \bibinfo{year}{2015}\natexlab{}.
\newblock \showarticletitle{FMTCP: A fountain code-based multipath transmission
  control protocol}.
\newblock \bibinfo{journal}{{\em IEEE/ACM Transactions on Networking (ToN)\/}}
  \bibinfo{volume}{23}, \bibinfo{number}{2} (\bibinfo{year}{2015}),
  \bibinfo{pages}{465--478}.
\newblock


\bibitem[\protect\citeauthoryear{De~Coninck and Bonaventure}{De~Coninck and
  Bonaventure}{2017}]%
        {mpquic}
\bibfield{author}{\bibinfo{person}{Quentin De~Coninck} {and}
  \bibinfo{person}{Olivier Bonaventure}.} \bibinfo{year}{2017}\natexlab{}.
\newblock \showarticletitle{Multipath QUIC: Design and Evaluation}. In
  \bibinfo{booktitle}{{\em 13th International Conference on emerging Networking
  EXperiments and Technologies (CoNEXT 2017).
  \url{http://multipath-quic.org}}}.
\newblock


\bibitem[\protect\citeauthoryear{Dong, Xu, Fu, and Cao}{Dong
  et~al\mbox{.}}{2017}]%
        {lamps}
\bibfield{author}{\bibinfo{person}{Enhuan Dong}, \bibinfo{person}{Mingwei Xu},
  \bibinfo{person}{Xiaoming Fu}, {and} \bibinfo{person}{Yu Cao}.}
  \bibinfo{year}{2017}\natexlab{}.
\newblock \showarticletitle{LAMPS: A Loss Aware Scheduler for Multipath TCP
  over Highly Lossy Networks}. In \bibinfo{booktitle}{{\em Local Computer
  Networks (LCN), 2017 IEEE 42nd Conference on}}. IEEE, \bibinfo{pages}{1--9}.
\newblock


\bibitem[\protect\citeauthoryear{Elliott}{Elliott}{1963}]%
        {elliottmodel}
\bibfield{author}{\bibinfo{person}{Edwin~O Elliott}.}
  \bibinfo{year}{1963}\natexlab{}.
\newblock \showarticletitle{Estimates of error rates for codes on burst-noise
  channels}.
\newblock \bibinfo{journal}{{\em The Bell System Technical Journal\/}}
  \bibinfo{volume}{42}, \bibinfo{number}{5} (\bibinfo{year}{1963}),
  \bibinfo{pages}{1977--1997}.
\newblock


\bibitem[\protect\citeauthoryear{et~al}{et~al}{2017}]%
        {quic-go}
\bibfield{author}{\bibinfo{person}{L.~Clemente et al}.}
  \bibinfo{year}{2017}\natexlab{}.
\newblock \bibinfo{title}{A QUIC implementation in pure go.}
  (\bibinfo{year}{2017}).
\newblock
\newblock
\shownote{\url{https://github.com/lucas-clemente/quic-go}.}


\bibitem[\protect\citeauthoryear{Ferlin, Kucera, Claussen, and Alay}{Ferlin
  et~al\mbox{.}}{[n. d.]}]%
        {mptcpfec}
\bibfield{author}{\bibinfo{person}{Simone Ferlin}, \bibinfo{person}{Stepan
  Kucera}, \bibinfo{person}{Holger Claussen}, {and} \bibinfo{person}{Özgü
  Alay}.} \bibinfo{year}{[n. d.]}\natexlab{}.
\newblock \showarticletitle{MPTCP meets FEC: Supporting Latency-Sensitive
  Applications over Heterogeneous Networks}.
\newblock \bibinfo{journal}{{\em IEEE/ACM Transactions on Networking\/}}
  (\bibinfo{year}{[n. d.]}).
\newblock


\bibitem[\protect\citeauthoryear{Fisher}{Fisher}{1949}]%
        {fisher1949design}
\bibfield{author}{\bibinfo{person}{Ronald~A Fisher}.}
  \bibinfo{year}{1949}\natexlab{}.
\newblock \showarticletitle{The design of experiments}.
\newblock  (\bibinfo{year}{1949}).
\newblock


\bibitem[\protect\citeauthoryear{Galanos, Peck, and Roca}{Galanos
  et~al\mbox{.}}{2011}]%
        {rtp-reedsolomon}
\bibfield{author}{\bibinfo{person}{Sarit Galanos}, \bibinfo{person}{Orly Peck},
  {and} \bibinfo{person}{Vincent Roca}.} \bibinfo{year}{2011}\natexlab{}.
\newblock \bibinfo{booktitle}{{\em RTP Payload Format for Reed Solomon FEC}}.
\newblock \bibinfo{type}{Internet-Draft}
  draft-galanos-fecframe-rtp-reedsolomon-03. \bibinfo{institution}{IETF
  Secretariat}.
\newblock
\showURL{%
\url{http://www.ietf.org/internet-drafts/draft-galanos-fecframe-rtp-reedsolomon-03.txt}}


\bibitem[\protect\citeauthoryear{Ha, Rhee, and Xu}{Ha et~al\mbox{.}}{2008}]%
        {cubic}
\bibfield{author}{\bibinfo{person}{Sangtae Ha}, \bibinfo{person}{Injong Rhee},
  {and} \bibinfo{person}{Lisong Xu}.} \bibinfo{year}{2008}\natexlab{}.
\newblock \showarticletitle{CUBIC: a new TCP-friendly high-speed TCP variant}.
\newblock \bibinfo{journal}{{\em ACM SIGOPS operating systems review\/}}
  \bibinfo{volume}{42}, \bibinfo{number}{5} (\bibinfo{year}{2008}),
  \bibinfo{pages}{64--74}.
\newblock


\bibitem[\protect\citeauthoryear{Hamilton, Iyengar, Swett, and Wilk}{Hamilton
  et~al\mbox{.}}{2016}]%
        {quic-draft-00}
\bibfield{author}{\bibinfo{person}{Ryan Hamilton}, \bibinfo{person}{Janardhan
  Iyengar}, \bibinfo{person}{Ian Swett}, {and} \bibinfo{person}{Alyssa Wilk}.}
  \bibinfo{year}{2016}\natexlab{}.
\newblock \bibinfo{booktitle}{{\em QUIC: A UDP-Based Secure and Reliable
  Transport for HTTP/2}}.
\newblock \bibinfo{type}{Internet-Draft}
  draft-hamilton-early-deployment-quic-00. \bibinfo{institution}{IETF
  Secretariat}.
\newblock
\showURL{%
\url{http://www.ietf.org/internet-drafts/draft-hamilton-early-deployment-quic-00.txt}}


\bibitem[\protect\citeauthoryear{Hesmans, Duchene, Paasch, Detal, and
  Bonaventure}{Hesmans et~al\mbox{.}}{2013}]%
        {tcpmiddlebox}
\bibfield{author}{\bibinfo{person}{Benjamin Hesmans}, \bibinfo{person}{Fabien
  Duchene}, \bibinfo{person}{Christoph Paasch}, \bibinfo{person}{Gregory
  Detal}, {and} \bibinfo{person}{Olivier Bonaventure}.}
  \bibinfo{year}{2013}\natexlab{}.
\newblock \showarticletitle{Are TCP extensions middlebox-proof?}. In
  \bibinfo{booktitle}{{\em Proceedings of the 2013 workshop on Hot topics in
  middleboxes and network function virtualization}}. ACM,
  \bibinfo{pages}{37--42}.
\newblock


\bibitem[\protect\citeauthoryear{Honda, Nishida, Raiciu, Greenhalgh, Handley,
  and Tokuda}{Honda et~al\mbox{.}}{2011}]%
        {honda2011still}
\bibfield{author}{\bibinfo{person}{Michio Honda}, \bibinfo{person}{Yoshifumi
  Nishida}, \bibinfo{person}{Costin Raiciu}, \bibinfo{person}{Adam Greenhalgh},
  \bibinfo{person}{Mark Handley}, {and} \bibinfo{person}{Hideyuki Tokuda}.}
  \bibinfo{year}{2011}\natexlab{}.
\newblock \showarticletitle{Is it still possible to extend TCP?}. In
  \bibinfo{booktitle}{{\em Proceedings of the 2011 ACM SIGCOMM conference on
  Internet measurement conference}}. ACM, \bibinfo{pages}{181--194}.
\newblock


\bibitem[\protect\citeauthoryear{Huitema, Shore, Mankin, Dickinson, and
  Iyengar}{Huitema et~al\mbox{.}}{2018}]%
        {dnsoquic}
\bibfield{author}{\bibinfo{person}{Christian Huitema}, \bibinfo{person}{Melinda
  Shore}, \bibinfo{person}{Allison Mankin}, \bibinfo{person}{Sara Dickinson},
  {and} \bibinfo{person}{Jana Iyengar}.} \bibinfo{year}{2018}\natexlab{}.
\newblock \bibinfo{booktitle}{{\em Specification of DNS over Dedicated QUIC
  Connections}}.
\newblock \bibinfo{type}{Internet-Draft} draft-huitema-quic-dnsoquic-03.
  \bibinfo{institution}{IETF Secretariat}.
\newblock
\showURL{%
\url{http://www.ietf.org/internet-drafts/draft-huitema-quic-dnsoquic-03.txt}}


\bibitem[\protect\citeauthoryear{Ian~Swett and Roca}{Ian~Swett and
  Roca}{2018}]%
        {coding-for-quic}
\bibfield{author}{\bibinfo{person}{Marie-Jose~Montpetit Ian~Swett} {and}
  \bibinfo{person}{Vincent Roca}.} \bibinfo{year}{2018}\natexlab{}.
\newblock \bibinfo{booktitle}{{\em Coding for QUIC}}.
\newblock \bibinfo{type}{Internet-Draft} draft-swett-nwcrg-coding-for-quic-00.
  \bibinfo{institution}{IETF Secretariat}.
\newblock
\showURL{%
\url{http://www.ietf.org/internet-drafts/draft-swett-nwcrg-coding-for-quic-00.txt}}


\bibitem[\protect\citeauthoryear{Iyengar and Sweet}{Iyengar and Sweet}{2017}]%
        {quic-rec-draft-07}
\bibfield{author}{\bibinfo{person}{Jana Iyengar} {and} \bibinfo{person}{Ian
  Sweet}.} \bibinfo{year}{2017}\natexlab{}.
\newblock \bibinfo{booktitle}{{\em QUIC Loss Detection and Congestion
  Control}}.
\newblock \bibinfo{type}{Internet-Draft} draft-ietf-quic-recovery-07.
  \bibinfo{institution}{IETF Secretariat}.
\newblock
\showURL{%
\url{https://www.ietf.org/archive/id/draft-ietf-quic-recovery-07.txt}}


\bibitem[\protect\citeauthoryear{Iyengar and Thomson}{Iyengar and
  Thomson}{2018}]%
        {quic-draft-11}
\bibfield{author}{\bibinfo{person}{Jana Iyengar} {and} \bibinfo{person}{Martin
  Thomson}.} \bibinfo{year}{2018}\natexlab{}.
\newblock \bibinfo{booktitle}{{\em QUIC: A UDP-Based Multiplexed and Secure
  Transport}}.
\newblock \bibinfo{type}{Internet-Draft} draft-ietf-quic-transport-11.
  \bibinfo{institution}{IETF Secretariat}.
\newblock
\showURL{%
\url{http://www.ietf.org/internet-drafts/draft-ietf-quic-transport-11.txt}}


\bibitem[\protect\citeauthoryear{Jacobson, Frederick, Casner, and
  Schulzrinne}{Jacobson et~al\mbox{.}}{2003}]%
        {rtp}
\bibfield{author}{\bibinfo{person}{Van Jacobson}, \bibinfo{person}{Ron
  Frederick}, \bibinfo{person}{Steve Casner}, {and} \bibinfo{person}{H
  Schulzrinne}.} \bibinfo{year}{2003}\natexlab{}.
\newblock \showarticletitle{RFC3550: RTP: A transport protocol for real-time
  applications}.
\newblock  (\bibinfo{year}{2003}).
\newblock


\bibitem[\protect\citeauthoryear{Kakhki, Jero, Choffnes, Nita-Rotaru, and
  Mislove}{Kakhki et~al\mbox{.}}{2017}]%
        {takinglonglookatquic}
\bibfield{author}{\bibinfo{person}{Arash~Molavi Kakhki},
  \bibinfo{person}{Samuel Jero}, \bibinfo{person}{David Choffnes},
  \bibinfo{person}{Cristina Nita-Rotaru}, {and} \bibinfo{person}{Alan
  Mislove}.} \bibinfo{year}{2017}\natexlab{}.
\newblock \showarticletitle{Taking a long look at QUIC: an approach for
  rigorous evaluation of rapidly evolving transport protocols}. In
  \bibinfo{booktitle}{{\em Proceedings of the 2017 Internet Measurement
  Conference}}. ACM, \bibinfo{pages}{290--303}.
\newblock


\bibitem[\protect\citeauthoryear{Katti, Rahul, Hu, Katabi, M{\'e}dard, and
  Crowcroft}{Katti et~al\mbox{.}}{2008}]%
        {katti2008xors}
\bibfield{author}{\bibinfo{person}{Sachin Katti}, \bibinfo{person}{Hariharan
  Rahul}, \bibinfo{person}{Wenjun Hu}, \bibinfo{person}{Dina Katabi},
  \bibinfo{person}{Muriel M{\'e}dard}, {and} \bibinfo{person}{Jon Crowcroft}.}
  \bibinfo{year}{2008}\natexlab{}.
\newblock \showarticletitle{XORs in the air: practical wireless network
  coding}.
\newblock \bibinfo{journal}{{\em IEEE/ACM Transactions on Networking (ToN)\/}}
  \bibinfo{volume}{16}, \bibinfo{number}{3} (\bibinfo{year}{2008}),
  \bibinfo{pages}{497--510}.
\newblock


\bibitem[\protect\citeauthoryear{Khalili, Gast, Popovic, Upadhyay, and
  Le~Boudec}{Khalili et~al\mbox{.}}{2012}]%
        {olia}
\bibfield{author}{\bibinfo{person}{Ramin Khalili}, \bibinfo{person}{Nicolas
  Gast}, \bibinfo{person}{Miroslav Popovic}, \bibinfo{person}{Utkarsh
  Upadhyay}, {and} \bibinfo{person}{Jean-Yves Le~Boudec}.}
  \bibinfo{year}{2012}\natexlab{}.
\newblock \showarticletitle{MPTCP is not pareto-optimal: performance issues and
  a possible solution}. In \bibinfo{booktitle}{{\em Proceedings of the 8th
  international conference on Emerging networking experiments and
  technologies}}. ACM, \bibinfo{pages}{1--12}.
\newblock


\bibitem[\protect\citeauthoryear{Langley, Riddoch, Wilk, Vicente, Krasic,
  Zhang, Yang, Kouranov, Swett, Iyengar, et~al\mbox{.}}{Langley
  et~al\mbox{.}}{2017}]%
        {langley2017quic}
\bibfield{author}{\bibinfo{person}{Adam Langley}, \bibinfo{person}{Alistair
  Riddoch}, \bibinfo{person}{Alyssa Wilk}, \bibinfo{person}{Antonio Vicente},
  \bibinfo{person}{Charles Krasic}, \bibinfo{person}{Dan Zhang},
  \bibinfo{person}{Fan Yang}, \bibinfo{person}{Fedor Kouranov},
  \bibinfo{person}{Ian Swett}, \bibinfo{person}{Janardhan Iyengar},
  {et~al\mbox{.}}} \bibinfo{year}{2017}\natexlab{}.
\newblock \showarticletitle{The QUIC transport protocol: Design and
  Internet-scale deployment}. In \bibinfo{booktitle}{{\em Proceedings of the
  Conference of the ACM Special Interest Group on Data Communication}}. ACM,
  \bibinfo{pages}{183--196}.
\newblock


\bibitem[\protect\citeauthoryear{Megyesi, Kr{\"a}mer, and Moln{\'a}r}{Megyesi
  et~al\mbox{.}}{2016}]%
        {howquicisquic}
\bibfield{author}{\bibinfo{person}{P{\'e}ter Megyesi}, \bibinfo{person}{Zsolt
  Kr{\"a}mer}, {and} \bibinfo{person}{S{\'a}ndor Moln{\'a}r}.}
  \bibinfo{year}{2016}\natexlab{}.
\newblock \showarticletitle{How quick is QUIC?}. In \bibinfo{booktitle}{{\em
  Communications (ICC), 2016 IEEE International Conference on}}. IEEE,
  \bibinfo{pages}{1--6}.
\newblock


\bibitem[\protect\citeauthoryear{Paasch, Khalili, and Bonaventure}{Paasch
  et~al\mbox{.}}{2013}]%
        {experimentaldesignmptcp}
\bibfield{author}{\bibinfo{person}{Christoph Paasch}, \bibinfo{person}{Ramin
  Khalili}, {and} \bibinfo{person}{Olivier Bonaventure}.}
  \bibinfo{year}{2013}\natexlab{}.
\newblock \showarticletitle{On the benefits of applying experimental design to
  improve multipath TCP}. In \bibinfo{booktitle}{{\em Proceedings of the ninth
  ACM conference on Emerging networking experiments and technologies}}. ACM,
  \bibinfo{pages}{393--398}.
\newblock


\bibitem[\protect\citeauthoryear{Park and Miller}{Park and Miller}{1988}]%
        {parkmiller}
\bibfield{author}{\bibinfo{person}{Stephen~K. Park} {and}
  \bibinfo{person}{Keith~W. Miller}.} \bibinfo{year}{1988}\natexlab{}.
\newblock \showarticletitle{Random number generators: good ones are hard to
  find}.
\newblock \bibinfo{journal}{{\it Commun. ACM}} \bibinfo{volume}{31},
  \bibinfo{number}{10} (\bibinfo{year}{1988}), \bibinfo{pages}{1192--1201}.
\newblock


\bibitem[\protect\citeauthoryear{Perkins}{Perkins}{2010}]%
        {rtpcongestion}
\bibfield{author}{\bibinfo{person}{Colin Perkins}.}
  \bibinfo{year}{2010}\natexlab{}.
\newblock \showarticletitle{RTP and the Datagram Congestion Control Protocol
  (DCCP)}.
\newblock  (\bibinfo{year}{2010}).
\newblock


\bibitem[\protect\citeauthoryear{Perkins, Hodson, and Hardman}{Perkins
  et~al\mbox{.}}{1998}]%
        {perkins1998survey}
\bibfield{author}{\bibinfo{person}{Colin Perkins}, \bibinfo{person}{Orion
  Hodson}, {and} \bibinfo{person}{Vicky Hardman}.}
  \bibinfo{year}{1998}\natexlab{}.
\newblock \showarticletitle{A survey of packet loss recovery techniques for
  streaming audio}.
\newblock \bibinfo{journal}{{\em IEEE Network\/}} \bibinfo{volume}{12},
  \bibinfo{number}{5} (\bibinfo{year}{1998}), \bibinfo{pages}{40--48}.
\newblock


\bibitem[\protect\citeauthoryear{Post}{Post}{2017}]%
        {Post_RS}
\bibfield{author}{\bibinfo{person}{Klaus Post}.}
  \bibinfo{year}{2017}\natexlab{}.
\newblock \bibinfo{title}{Reed-Solomon Erasure Coding in Go}.
  (\bibinfo{year}{2017}).
\newblock
\newblock
\shownote{\url{https://github.com/klauspost/reedsolomon}.}


\bibitem[\protect\citeauthoryear{Raiciu, Paasch, Barre, Ford, Honda, Duchene,
  Bonaventure, and Handley}{Raiciu et~al\mbox{.}}{2012}]%
        {deployingmptcp}
\bibfield{author}{\bibinfo{person}{Costin Raiciu}, \bibinfo{person}{Christoph
  Paasch}, \bibinfo{person}{Sebastien Barre}, \bibinfo{person}{Alan Ford},
  \bibinfo{person}{Michio Honda}, \bibinfo{person}{Fabien Duchene},
  \bibinfo{person}{Olivier Bonaventure}, {and} \bibinfo{person}{Mark Handley}.}
  \bibinfo{year}{2012}\natexlab{}.
\newblock \showarticletitle{How hard can it be? designing and implementing a
  deployable multipath TCP}. In \bibinfo{booktitle}{{\em Proceedings of the 9th
  USENIX conference on Networked Systems Design and Implementation}}. USENIX
  Association, \bibinfo{pages}{29--29}.
\newblock


\bibitem[\protect\citeauthoryear{Reed and Solomon}{Reed and Solomon}{1960}]%
        {reedsolomoncodes}
\bibfield{author}{\bibinfo{person}{Irving~S Reed} {and}
  \bibinfo{person}{Gustave Solomon}.} \bibinfo{year}{1960}\natexlab{}.
\newblock \showarticletitle{Polynomial codes over certain finite fields}.
\newblock \bibinfo{journal}{{\em Journal of the society for industrial and
  applied mathematics\/}} \bibinfo{volume}{8}, \bibinfo{number}{2}
  (\bibinfo{year}{1960}), \bibinfo{pages}{300--304}.
\newblock


\bibitem[\protect\citeauthoryear{Roca and Cunche}{Roca and Cunche}{2012}]%
        {rfc6816}
\bibfield{author}{\bibinfo{person}{V Roca} {and} \bibinfo{person}{M Cunche}.}
  \bibinfo{year}{2012}\natexlab{}.
\newblock \bibinfo{booktitle}{{\em J. Lacan," Simple Low-Density Parity Check
  (LDPC) Staircase Forward Error Correction (FEC) Scheme for FECFRAME}}.
\newblock \bibinfo{type}{{T}echnical {R}eport}. \bibinfo{institution}{RFC 6816,
  DOI 10.17487/RFC6816, December 2012,< http://www. rfc-editor.
  org/info/rfc6816}.
\newblock


\bibitem[\protect\citeauthoryear{Roca and Teibi}{Roca and Teibi}{2018a}]%
        {rlc-fecframe}
\bibfield{author}{\bibinfo{person}{Vincent Roca} {and}
  \bibinfo{person}{Belkacem Teibi}.} \bibinfo{year}{2018}\natexlab{a}.
\newblock \bibinfo{booktitle}{{\em Sliding Window Random Linear Code (RLC)
  Forward Erasure Correction (FEC) Schemes for FECFRAME}}.
\newblock \bibinfo{type}{Internet-Draft} draft-ietf-tsvwg-rlc-fec-scheme-02.
  \bibinfo{institution}{IETF Secretariat}.
\newblock
\showURL{%
\url{http://www.ietf.org/internet-drafts/draft-ietf-tsvwg-rlc-fec-scheme-02.txt}}


\bibitem[\protect\citeauthoryear{Roca and Teibi}{Roca and Teibi}{2018b}]%
        {rlc-fecframe-05}
\bibfield{author}{\bibinfo{person}{Vincent Roca} {and}
  \bibinfo{person}{Belkacem Teibi}.} \bibinfo{year}{2018}\natexlab{b}.
\newblock \bibinfo{booktitle}{{\em Sliding Window Random Linear Code (RLC)
  Forward Erasure Correction (FEC) Schemes for FECFRAME}}.
\newblock \bibinfo{type}{Internet-Draft} draft-ietf-tsvwg-rlc-fec-scheme-05.
  \bibinfo{institution}{IETF Secretariat}.
\newblock
\showURL{%
\url{http://www.ietf.org/internet-drafts/draft-ietf-tsvwg-rlc-fec-scheme-05.txt}}


\bibitem[\protect\citeauthoryear{Roca, Teibi, Burdinat, Tran, and Thienot}{Roca
  et~al\mbox{.}}{2017}]%
        {roca2017less}
\bibfield{author}{\bibinfo{person}{Vincent Roca}, \bibinfo{person}{Belkacem
  Teibi}, \bibinfo{person}{Christophe Burdinat}, \bibinfo{person}{Tuan Tran},
  {and} \bibinfo{person}{C{\'e}dric Thienot}.} \bibinfo{year}{2017}\natexlab{}.
\newblock \showarticletitle{Less latency and better protection with AL-FEC
  sliding window codes: A robust multimedia CBR broadcast case study}. In
  \bibinfo{booktitle}{{\em Wireless and Mobile Computing, Networking and
  Communications (WiMob),}}. IEEE, \bibinfo{pages}{1--8}.
\newblock


\bibitem[\protect\citeauthoryear{Roca, Watson, and Begen}{Roca
  et~al\mbox{.}}{2011}]%
        {fecframe}
\bibfield{author}{\bibinfo{person}{Vincent Roca}, \bibinfo{person}{Mark
  Watson}, {and} \bibinfo{person}{Ali~C. Begen}.}
  \bibinfo{year}{2011}\natexlab{}.
\newblock \bibinfo{title}{{Forward Error Correction (FEC) Framework}}.
\newblock \bibinfo{howpublished}{RFC 6363}.   (\bibinfo{date}{Oct.}
  \bibinfo{year}{2011}).
\newblock
\showDOI{%
\url{https://doi.org/10.17487/RFC6363}}


\bibitem[\protect\citeauthoryear{Santiago, Claeys-Bruno, and Sergent}{Santiago
  et~al\mbox{.}}{2012}]%
        {wsp}
\bibfield{author}{\bibinfo{person}{J Santiago}, \bibinfo{person}{M
  Claeys-Bruno}, {and} \bibinfo{person}{M Sergent}.}
  \bibinfo{year}{2012}\natexlab{}.
\newblock \showarticletitle{Construction of space-filling designs using WSP
  algorithm for high dimensional spaces}.
\newblock \bibinfo{journal}{{\em Chemometrics and Intelligent Laboratory
  Systems\/}}  \bibinfo{volume}{113} (\bibinfo{year}{2012}),
  \bibinfo{pages}{26--31}.
\newblock


\bibitem[\protect\citeauthoryear{Sundararajan, Shah, M{\'e}dard, Jakubczak,
  Mitzenmacher, and Barros}{Sundararajan et~al\mbox{.}}{2011}]%
        {sundararajan2011network}
\bibfield{author}{\bibinfo{person}{Jay~Kumar Sundararajan},
  \bibinfo{person}{Devavrat Shah}, \bibinfo{person}{Muriel M{\'e}dard},
  \bibinfo{person}{Szymon Jakubczak}, \bibinfo{person}{Michael Mitzenmacher},
  {and} \bibinfo{person}{Joao Barros}.} \bibinfo{year}{2011}\natexlab{}.
\newblock \showarticletitle{Network coding meets TCP: Theory and
  implementation}.
\newblock \bibinfo{journal}{{\it Proc. IEEE}} \bibinfo{volume}{99},
  \bibinfo{number}{3} (\bibinfo{year}{2011}), \bibinfo{pages}{490--512}.
\newblock


\bibitem[\protect\citeauthoryear{Swett}{Swett}{2016}]%
        {Swett_FEC:2016}
\bibfield{author}{\bibinfo{person}{I. Swett}.} \bibinfo{year}{2016}\natexlab{}.
\newblock \bibinfo{title}{{QUIC FEC v1}}.  (\bibinfo{date}{Feb}
  \bibinfo{year}{2016}).
\newblock
\newblock
\shownote{unpublished draft,
  \url{https://docs.google.com/document/d/1Hg1SaLEl6T4rEU9j-isovCo8VEjjnuCPTcLNJewj7Nk/edit}.}


\bibitem[\protect\citeauthoryear{Tournoux, Lochin, Lacan, Bouabdallah, and
  Roca}{Tournoux et~al\mbox{.}}{2011}]%
        {tetrys}
\bibfield{author}{\bibinfo{person}{Pierre~Ugo Tournoux},
  \bibinfo{person}{Emmanuel Lochin}, \bibinfo{person}{J{\'e}r{\^o}me Lacan},
  \bibinfo{person}{Amine Bouabdallah}, {and} \bibinfo{person}{Vincent Roca}.}
  \bibinfo{year}{2011}\natexlab{}.
\newblock \showarticletitle{On-the-fly erasure coding for real-time video
  applications}.
\newblock \bibinfo{journal}{{\em IEEE Transactions on Multimedia\/}}
  \bibinfo{volume}{13}, \bibinfo{number}{4} (\bibinfo{year}{2011}),
  \bibinfo{pages}{797--812}.
\newblock


\bibitem[\protect\citeauthoryear{Zhang, Ng, Nandi, Riedi, Druschel, and
  Wang}{Zhang et~al\mbox{.}}{2006}]%
        {zhang2006measurement}
\bibfield{author}{\bibinfo{person}{Bo Zhang}, \bibinfo{person}{TS~Eugene Ng},
  \bibinfo{person}{Animesh Nandi}, \bibinfo{person}{Rudolf Riedi},
  \bibinfo{person}{Peter Druschel}, {and} \bibinfo{person}{Guohui Wang}.}
  \bibinfo{year}{2006}\natexlab{}.
\newblock \showarticletitle{Measurement-Based Analysis, Modeling, and Synthesis
  of the Internet Delay Space}.
\newblock  (\bibinfo{year}{2006}).
\newblock


\end{thebibliography}
